\documentclass{aa}

\usepackage{graphicx}
\usepackage{txfonts}
\usepackage{xcolor}
\usepackage{cprotect}
\usepackage{pdflscape}
\usepackage[switch]{lineno}
\usepackage{linenoaa}
\usepackage{multicol}
\usepackage[breaklinks]{hyperref}
\usepackage{lastpage}
\RequirePackage{etex}

\bibpunct{(}{)}{;}{a}{}{,}             
\definecolor{cobalt}{rgb}{0.06, 0.2, 0.65}
\hypersetup{
  colorlinks,
  citecolor=cobalt,
  linkcolor=[rgb]{0.8, 0.2, 1.0},
  urlcolor=cobalt,
}

\makeatletter
\newcommand\DefineObj[3][\@empty]{%
  \expandafter\newcommand\csname pkgwobj@l#2\endcsname{#3}%
  \ifx\@empty#1%
    \expandafter\newcommand\csname pkgwobj@s#2\endcsname{#3}%
  \else%
    \expandafter\newcommand\csname pkgwobj@s#2\endcsname{#1}%
  \fi}%
\newcommand{\pkgw@printobj@long}[1]{%
  \expandafter\ifx\csname pkgwobj@l#1\endcsname\relax%
    \textbf{[unknown object!]}%
  \else%
    \csname pkgwobj@l#1\endcsname%
  \fi}%
\newcommand{\pkgw@printobj@short}[1]{%
  \expandafter\ifx\csname pkgwobj@l#1\endcsname\relax%
    \textbf{[unknown object!]}%
  \else%
    \csname pkgwobj@s#1\endcsname%
  \fi}%
\newcommand{\obj}{\@ifstar{\pkgw@printobj@long}{\pkgw@printobj@short}}%
\makeatother

\DefineObj[2M2206$-$42]{2206}{2MASSW J2206450$-$421721}
\DefineObj[2M0355$+$11]{0355}{2MASS J03552337$+$1133437}
\DefineObj[W0047$+$68]{0047}{2MASS J00470038$+$6803543}
\DefineObj[2M2244$+$20]{2244}{2MASS J22443167$+$2043433}
\DefineObj[2M0642$+$41]{0642}{2MASS J06420559$+$4101599}
\DefineObj[W1741$-$46]{1741}{2MASS J17410280$-$4642218}
\DefineObj{h2o}{H$_2$O}
\DefineObj{ch4}{CH$_4$}
\DefineObj{nh3}{NH$_3$}
\DefineObj{co2}{CO$_2$}
\DefineObj{enstatite}{MgSiO$_3$}
\DefineObj{forsterite}{Mg$_2$SiO$_4$}
\DefineObj{quartz}{SiO$_2$}

\begin{document}

   \title{Clouds with a silicate lining: Using JWST spectra to probe atmospheric diversity in young AB Dor L dwarfs}
   \titlerunning{Using JWST spectra to probe atmospheric diversity in young AB Dor L dwarfs}

   \author{M.~B.~Lam
          \inst{1}
          \and J.~M. ~Vos \inst{1}
          \and G.~Suárez \inst{2}
          \and C.-C.~Hsu \inst{3}
          \and T.~P.~Bickle \inst{4}
          \and J.~Faherty \inst{2}
          \and J.~Gagné \inst{5,6}
          \and D.~Bardalez~Gagliuffi \inst{7}
          \and B.~Biller \inst{8}
          \and B.~Burningham \inst{9}
          \and K.~L.~Cruz \inst{2,10,15}
          \and C.~V.~Morley \inst{11}
          \and S.~Luszcz-Cook \inst{2,12}
          \and S.~Lawsky \inst{13}
          \and C.~L.~Phillips \inst{14}
          \and A.~Rothermich \inst{2,10,15}
          }

    \institute{School of Physics, Trinity College Dublin, The University of Dublin, Dublin 2, Ireland\\
    \email{lamm3@tcd.ie}
    \and
    Department of Astrophysics, American Museum of Natural History, Central Park West at 79th St, New York, NY 10024, USA
    \and
    Center for Interdisciplinary Exploration and Research in Astrophysics (CIERA), Northwestern University, 1800 Sherman, Evanston, IL 60201, USA
    \and School of Physical Sciences, The Open University, Milton Keynes, MK7 6AA, UK
    \and
    Plan\'etarium de Montr\'eal, Espace pour la Vie, 4801 av. Pierre-de Coubertin, Montr\'eal, Qu\'ebec, Canada
    \and
    Trottier Institute for Research on Exoplanets, Universit\'e de Montr\'eal, D\'epartement de Physique, C.P.~6128 Succ. Centre-ville, Montr\'eal, QC H3C~3J7, Canada
    \and
    Department of Physics \& Astronomy, Amherst College, 25 East Dr., Amherst, MA 01002, USA
    \and
    Institute for Astronomy, University of Edinburgh, Royal Observatory, Edinburgh EH9 3HJ
    \and Centre for Astrophysics Research, Department of Physics, Astronomy and Mathematics, University of Hertfordshire, Hatfield AL10 9AB, UK
    \and Department of Physics and Astronomy, Hunter College, City University of New York, 695 Park Avenue, New York, NY, 10065, USA
    \and Department of Astronomy, University of Texas at Austin, 2515 Speedway, Austin, TX 78712, USA
    \and New York University, New York, NY 10003, USA
    \and
    Department of Astronomy, Columbia University, 550 West 120th Street, New York, NY 10027, USA
    \and
    Department of Astronomy \& Astrophysics, University of California, Santa Cruz, CA 95064, USA
    \and Department of Physics, Graduate Center, City University of New York, 365 5th Avenue, New York, NY 10016, USA
    }

   \date{}

  \abstract{}
  {We present the first full JWST NIRSpec Prism and MIRI LRS 0.6 -- 14 $\mu$m ($R\sim 100$) spectra and analysis of five $\sim$133\,Myr L dwarf members of the AB Doradus moving group and one probable $\sim500$\,Myr T dwarf of the Oceanus moving group with known inclination angles between $\sim 23$ -- 90$^{\circ}$: \protect\obj*{0047}, \protect\obj*{0355}, \protect\obj*{0642}, \protect\obj*{1741}, \protect\obj*{2206}, and \protect\obj*{2244}.}
   {We construct near-complete spectral energy distributions of each of our objects to measure their bolometric luminosities, and estimate their fundamental parameters ($T_{\text{eff}}$, radius, $M$ and $\log g$). We use cross-sections of relevant gases to identify the species that are present in each atmosphere. Of particular interest is the silicate absorption feature at 8 -- 11 $\mu$m, which provides insight into the complex cloud structure of brown dwarfs. We examine this silicate absorption feature in detail and also test whether there exists a latitudinal dependence in the silicate absorption feature within a coeval sample of brown dwarfs.}
   {Various molecular absorption bands are visible in our spectra, including \protect\obj{h2o}, \protect\obj{ch4}, CO and \protect\obj{co2}. The shape of the silicate absorption feature varies within our sample, and we find that 4/5 of our L type objects agree with previously observed trends stating that objects viewed equator-on have deeper silicate absorption. We highlight \protect\obj{1741} as an outlier in our sample with an unusually strong silicate absorption given its near pole-on orientation. We also present a tentative correlation between the wavelength of peak silicate absorption and inclination, which may suggest variations in cloud chemical composition or physical properties.}
   {We find an unexpected spectral diversity within our sample, which motivates future studies on these objects through atmospheric retrievals, which will determine the silicate cloud composition and reveal whether there exists a trend with inclination.}
   \keywords{brown dwarfs; Planets and satellites: atmospheres;
               }

   \maketitle
   \nolinenumbers

\section{Introduction}
Brown dwarfs are substellar objects that exist in the intermediate mass range ($M\sim 13$ -- {78.5} $M_{\text{Jup}}$) {\citep{Kumar1963,Chabrier2023}} between stars and planets. Brown dwarfs cool over time, reaching temperatures that rival solar system objects or the coldest directly imaged exoplanets (e.g. WISE 0855$-$0714 at 250 K \citep{Luhman2014}, Eps~Ind~Ab at 275 K \citep{Matthews2024} and 14~Her~c at 250 -- 300 K \citep{BGagliuffi2021,BGagliuffi2025}). They can be considered "overachieving planets" as they are similar to giant planets in terms of their temperature, composition and radius \citep{Burrows2001}, but tend to be more massive. Direct imaging of exoplanet atmospheres is challenging due to their bright host star drowning out the light emitted from the exoplanet itself. Brown dwarfs provide an alternative route to study extrasolar atmospheres, as they are often found in isolation, alleviating the problem of blocking a bright host star. As a result, we have a larger sample of brown dwarfs and higher quality observations compared to exoplanets. Studying brown dwarfs in the context of planetary atmospheres will allow us to better understand exoplanet atmospheric processes and dynamics \citep{Apai2017,Showman2020}.

Over time, brown dwarfs cool and evolve across the L-T-Y spectral types, changing their infrared (IR) colours, which reflect changes in their temperature and atmospheric chemical composition. L spectral type brown dwarfs have similar effective temperatures ($T_{\text{eff}}$) to many directly imaged exoplanets \citep[eg. HR~8799~bcde;][]{Marois2008,Marois2010}. Thus, we can use L dwarfs as proxies for gas giants. L dwarfs are defined by the strengthening of neutral alkali lines, oxides and hydrides in their spectra, which can be translated to effective temperatures between 1300 -- 2500 K \citep{Kirkpatrick2005}. L dwarfs are of particular interest as silicate clouds condense in the photosphere, making these brown dwarfs appear redder in the near-infrared (NIR) compared to other spectral types \citep[e.g.][]{AckermanMarley2001,Marley2002,Tsuji2002,Knapp2004,Burrows2006,Cushing2006,Marley2010}. At the L-T spectral transition ($\sim 1000-1200$ K for low gravity objects with $\log g \sim 4-4.5$), the influence of silicate clouds on the spectral morphology reduces, shifting their NIR colours to bluer wavelengths \citep{AckermanMarley2001,SaumonMarley2008}. Alternative theories suggest that the brightening of the J-band for L dwarfs close to the L-T transition is not caused by clouds, but instead from energy transport from non-equilibrium carbon chemistry \citep{Tremblin2016}. These silicate clouds are thought to be one of the primary drivers of variability observed on L dwarfs in IR wavelengths \citep[e.g.][]{Artigau2009,Radigan2012,Apai2013,Lew2016,Vos2020,SuarezMetchev2022,Vos2022}. Studying cloud variability in brown dwarfs provides an enhanced insight into the atmospheric dynamics of extrasolar gas giant planets with similar fundamental properties \citep[e.g.][]{Ge2019,Bowler2020,Gao2021}.

Silicate clouds in L dwarfs are thought to be produced by Si- and Mg-chemistry forming three main compositions \citep{LoddersFegley2002}: enstatite (\obj{enstatite}), forsterite (\obj{forsterite}) and quartz (\obj{quartz}), which shape the silicate absorption feature present at mid-IR wavelengths \citep{LunaMorley2021}. This silicate absorption at 8 -- 11 $\mu$m caused by Si-O bonds stretching has previously been observed in many brown dwarfs by the \textit{Spitzer} Infrared Spectrograph (IRS) \citep{Cushing2006,Stephens2009,SuarezMetchev2022}. The silicate absorption feature has also been detected in brown dwarfs and directly imaged exoplanets with the Mid-Infrared Instrument (MIRI) on the James Webb Space Telescope (JWST) \citep[e.g. ][]{Miles2023,Biller2024,Chen2025,Hoch2025,Molliere2025}. The shape of the silicate feature (peak wavelength, width, depth) provides insight into the silicate composition and grain size \citep{LunaMorley2021,SuarezMetchev2023}. Atmospheric retrievals, a data-driven modelling technique, have been used to disentangle the cloudy structure of brown dwarf atmospheres, revealing structures like patchy \obj{forsterite} clouds \citep{Vos2023} or layers of \obj{enstatite} and \obj{quartz} \citep{Burningham2021}, supporting the hypothesis that clouds drive variability observed on brown dwarfs. Recently, \obj{quartz} clouds have also been observed in the atmospheres of transiting hot Jupiter exoplanets \citep{Grant2023} with JWST/MIRI.
%(e.g. VHS~1256~b \citep[][]{Miles2023}, WISE~1049B \citep[][]{Biller2024,Chen2025}, YSES-1~c \citep{Hoch2025} and PSO~J318 \citep{Molliere2025})

In recent years, the inclination angle, $i$, has emerged as an important property of an observed atmosphere, where brown dwarfs that are viewed equator-on appear both redder and more variable than those viewed pole-on \citep{Vos2017,Vos2018,Vos2020}. We define the inclination angle to be the viewing angle relative to the object's equator from our line of sight. To calculate the inclination angle of each object, we require high resolution spectra to measure the projected rotational velocity, $v\sin i$. Combined with the radius from fitting evolutionary models and rotational period obtained from variability studies, we can calculate the inclination angle. \citet{Suarez2023} also observed that the mid-IR silicate absorption feature was more prominent for brown dwarfs observed closer to equator-on. Thus, we can interpret that objects viewed equator-on appear redder and more variable due to varying temperature or particle size of silicate cloud layers. Higher altitude clouds are expected to be redder because they are cooler than clouds deeper in the atmosphere \citep{AckermanMarley2001}. This latitudinal variation in cloud thickness has also been theoretically predicted by General Circulation Model (GCM) simulations \citep{Showman2020,TanShowman2021}. The polar vortex hypothesis \citep{Fuda2024}, for example, explains how jet-dominated bands across the equator and vortex-dominated poles could cause similar observed variations. However, different ages or overall chemical compositions in previously studied samples could cause these latitudinal differences. In this work, we study the silicate feature for a sample of coeval objects for the first time and we investigate the hypothesis that atmospheric dynamics in L dwarfs drive equator-to-pole differences in cloud properties.

In this paper, we present the first analyses of JWST spectra of a sample of $\sim$133\,Myr AB Doradus moving group (hereafter, AB Dor) objects. Section~\ref{sec:sample-properties} describes relevant properties of our sample of AB Dor members. In Section~\ref{sec:observations} we discuss the observations and data reduction of our JWST spectra, as well as high-resolution spectra we obtained from IGRINS for $v\sin i$ measurements. In Section~\ref{sec:spectral-typing}, we present new and updated spectral types for each of our objects based on our new spectra. In Section~\ref{sec:fundamental-parameters}, we construct a full Spectral Energy Distribution (SED) for each of our objects to measure the bolometric luminosity, effective temperature, mass, radius and surface gravity. We also present new rotational and radial velocity measurements from the Immersion GRating INfrared Spectrometer (IGRINS) for two of the objects in our sample in Section~\ref{sec:igrins}. In Section~\ref{sec:key-molecular-absorptions}, we highlight which molecules are present in the atmospheres of the objects in our sample, and measure the water, methane and silicate indices of each object. We also discuss whether we observe any trends between inclination and the silicate feature. In Section~\ref{sec:spectral-comparisons}, we compare our JWST spectra to previous \textit{Spitzer} observations to test for evidence for long-term variability and JWST spectra of other similar planetary mass objects from the literature.

\section{Sample properties}
\label{sec:sample-properties}

Our sample comprises six well-characterised L dwarfs: \obj*{0047}, \obj*{0355}, \obj*{0642}, \obj*{1741}, \obj*{2206}, and \obj*{2244} (hereafter \obj{0047}, \obj{0355}, \obj{0642}, \obj{1741}, \obj{2206}, and \obj{2244}, respectively). We have chosen these targets because all except one are bona fide (BF) members of the AB Dor moving group (see Table~\ref{table:properties}). Objects in the same moving group form together from the same molecular cloud, and hence have the same age and similar chemical composition. Therefore, studying low-temperature objects in the same moving group provides an ideal coeval sample of exoplanet analogues, where we can uniquely isolate effects due to effective temperature or inclination. All targets except \obj{2206} have published variability detections and period measurements with \textit{Spitzer} \citep{Vos2018,Vos2022} and combined with their $v\sin i$ we can measure their inclinations. For \obj{2206}, we can estimate its inclination using the range of expected rotational periods for AB Dor objects. This makes our sample an ideal testbed for exploring potential trends with inclination. We also have the opportunity to characterise this unique sample in unprecedented detail with our high-quality observations.\\
\\
Using the kinematic properties of our targets, including parallax ($\pi$), proper motion ($\mu$) and radial velocity (RV), we calculated the membership probability for each target with the BANYAN $\Sigma$ tool \citep{banyansigma}, and present them in Table~\ref{table:properties}. We use the membership definitions from \citet{Faherty2016} to describe our objects. BF members have full kinematic information (proper motions and radial velocities) and a high probability (>90\%) of membership. Ambiguous members (AM) require more kinematic information because they could belong to more than one group. Since our sample are either BF or AM members of the young AB Dor moving group they are likely to be the same age \citep[$133^{+15}_{-20}$ Myr;][]{Gagne2018} and have similar metallicity.

\obj{0642} is the only AM member in our sample because it is missing an RV measurement, which could place it in the Oceanus moving group ($\sim 500$ Myr) instead \citep{Gagne2023}. With the original BANYAN $\Sigma$ models \citep{banyansigma}, we find a 74.4\% membership probability in AB Dor, with an optimal RV of $0.2\pm1.4$ km/s. With newer models \citep{Gagne2026}, we get a higher membership probability in the Oceanus moving group with 88.2\% and a predicted RV of $12.9\pm0.9$ km/s. Updated parallax and proper motion measurements from \citet{Best2024} favour the Oceanus moving group, however we require a precise RV measurement for \obj{0642} to determine which moving group it is a member of.\\
\\
\begin{table*}
\caption{Kinematic properties for objects in our sample.}\label{table:properties}
\centering
\begin{tabular}{lcccccccccc} 
\hline\hline    
\noalign{\smallskip}

Object  & $\pi$ & $\mu_\text{R.A.}$ & $\mu_\text{Dec.}$ & RV & BANYAN $\Sigma$ & Membership& Reference \\
 &mas & mas/yr &mas/yr&km/s&\%&\\
\hline
\noalign{\smallskip}
  \obj{0047}  & $82\pm3$   & $380.7\pm1.1$ & $-204.2\pm1.4$ & $-19.8^{+0.1}_{-0.2}$ &99.9 &BF& 4, 5, 8\\
  \obj{0355}  &  $109.1\pm0.5$ & $223.2\pm0.6$ & $-631.3\pm0.4$ & $11.92\pm0.22$ &99.9&BF& 1, 7\\
  \obj{0642}    &  $62.6\pm0.9$ & $-2.0\pm1.2$ & $-383.1\pm1.2$ & ... &74.4& AM* &2,3 \\
  \obj{1741}   &  $50.5\pm2.9$ & $-29.2\pm2.1$ & $356.5\pm2.1$ & $2.54\pm0.11$ & 99.9&BF&3, TW\\
  \obj{2206}  & $34.1\pm1.3$ & $128.7\pm0.9$ & $-184.9\pm0.9$ & $6.8^{+0.15}_{-0.16}$ & 99.8&BF&1, 4, TW\\
  \noalign{\smallskip}
  \obj{2244}  &  $54\pm4$ & $230.3\pm0.9$ & $-234.8\pm1.0$ & $-16.0^{+0.8}_{-0.9}$ &99.6&BF &1, 4, 6, 8\\
\noalign{\smallskip}
\hline
\end{tabular}
\tablefoot{BF: Bona fide, AM: Ambiguous member

*: \obj{0642} may be part of the $\sim 500$ Myr Oceanus moving group, but can only be confirmed with a RV measurement.}
\tablebib{(1)~\citet{Gaia2023}; (2) \cite{Best2021}; 
(3) \citet{Kirkpatrick2021}; (4) \citet{Liu2016}; (5) \citet{Gizis2015}; 
(6) \citet{Faherty2016}; (7) \citet{Blake2010}; (8) \citet{Vos2018}; TW: This Work
}
\end{table*}

\begin{figure}
    \centering
    \includegraphics[width=\linewidth]{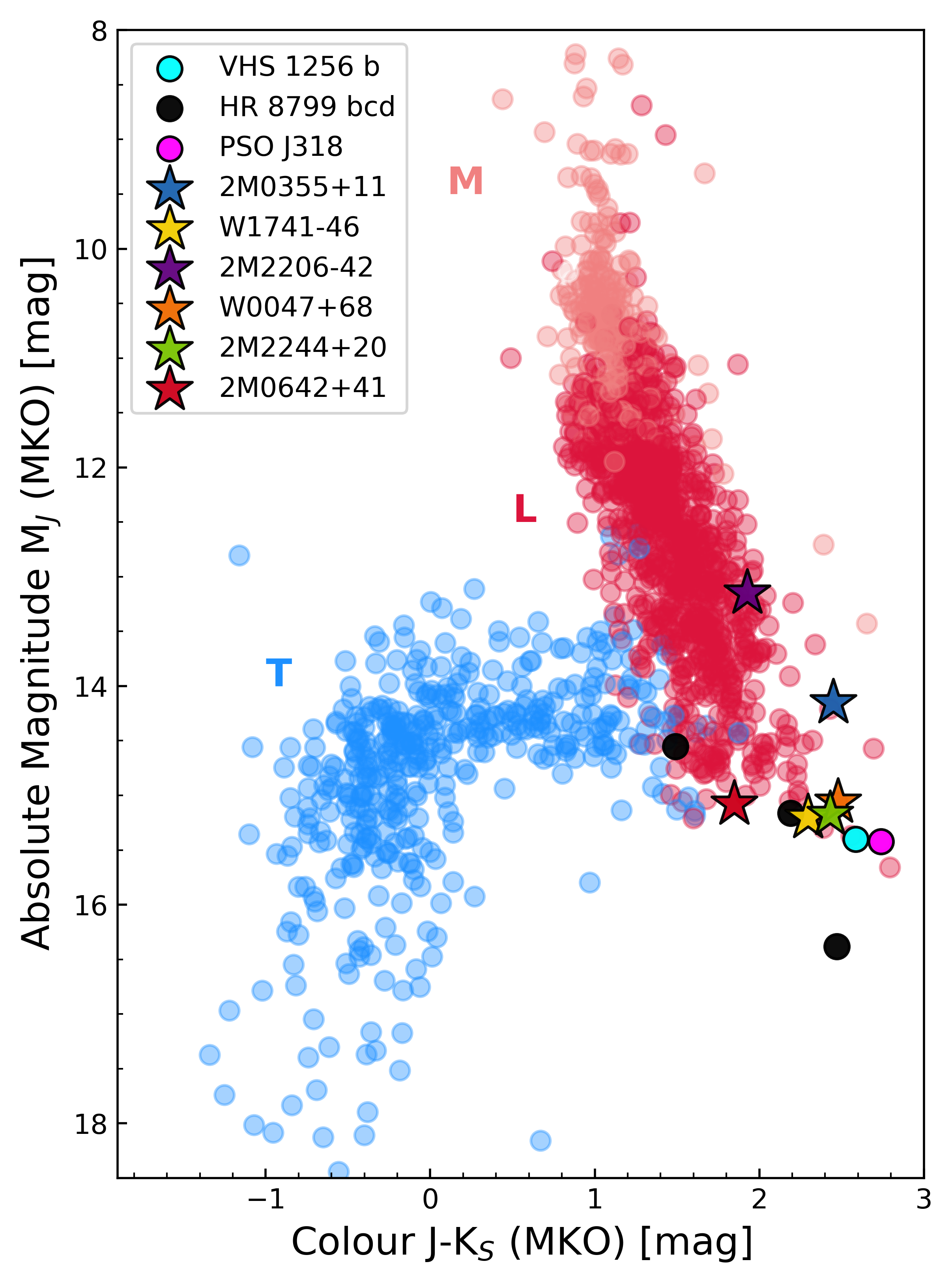}
    \caption{Colour-magnitude diagram indicating locations of our targets (coloured stars) at the L-T transition. The location of VHS~1256 b has also been included as a cyan circle. Planets in the HR~8799 system are also indicated as black circles. PSO~J318 is indicated as a pink circle. The coral points are M dwarfs, dark red points are L dwarfs, and blue points are T dwarfs. Photometric data for each point was obtained from the Ultracool Sheet \citep{UltraCoolSheet}.}
    \label{fig:colourmag}
\end{figure}

\noindent The location of each object on a NIR colour-magnitude diagram is shown in Fig.~\ref{fig:colourmag}. We focused on the J and K bands to highlight previous observations of the unusually red colours of objects in our sample. Our objects span the L dwarf to L-T transition range and are redder than most L dwarfs, a key property of young objects \citep{Faherty2016,Liu2016}. These redder colours are caused by the presence of silicate clouds in the photosphere \citep[e.g.][]{SaumonMarley2008,Faherty2013,Best2015,SuarezMetchev2022}. Our sample of brown dwarfs are comparable on the colour-magnitude diagram to VHS J125601.92-125723.9 b (hereafter VHS~1256~b) \citep{Gauza2015}, a planetary-mass companion that was observed by the first JWST early-release science program \citep{Miles2023}, a planetary-mass object PSO J318.5338-22.8603 (hereafter PSO J318) \citep{Liu2013}, and the HR 8799 planets \citep{Marois2008,Marois2010} (see Fig.~\ref{fig:colourmag}). Thus, our sample of brown dwarfs provides critical context to interpret observations of directly imaged exoplanets.\\
\\

\begin{table*}
\caption{Variability and inclination measurements for objects in our sample.}\label{table:variability}
\centering
\begin{tabular}{lccccccccc} 
\hline\hline 
\noalign{\smallskip}

Object & Period [hr]  & 3.6 Amplitude [\%] &Reference& $v\sin i$ [km/s]& $i$ [$\degr$]  &  Reference\\
\hline
\noalign{\smallskip}
  \obj{0047} & $16.4\pm0.2$ & $1.07\pm0.04$ & 2 & $9.8\pm0.3$ & $85^{+5}_{-9}$ & 2\\
  \obj{0355} & $9.53\pm 0.19$ & $0.26\pm0.02$ & 1 & $12.31\pm0.15$ & $50\pm2$ & 3, 4\\
  \obj{0642} & $10.11\pm0.06$ & $2.16\pm0.16$ & 1 &... &... &\\
  \obj{1741} & $15.00^{+0.71}_{-0.57}$ & $0.35\pm0.03$ & 1 &$3.88^{+0.46}_{-0.40}$& $23.6 ^{+3.4}_{-3.6}$&TW\\
  \noalign{\smallskip}
  \obj{2206} & {$9 - 17$*} & $< 0.33$ & 1 &$12.49^{+0.40}_{-0.28}$&48.6 -- 90{*}& TW\\
  \noalign{\smallskip}
  \obj{2244} & $11.0 \pm 2.0$ & $0.8\pm0.2$ & 2 & $14.3^{+1.3}_{-1.5}$ & $76^{+14}_{-20}$ & 2\\
\noalign{\smallskip}
\hline
\end{tabular}
\tablefoot{{*: \obj{2206} has no detected variability, so these estimates are based on the range of expected rotation periods for AB Dor members, 9 -- 17 h \citep{Vos2022} and its measured $v\sin i$.}}
\tablebib{TW: This work; (1) \citet{Vos2022}; (2) \citet{Vos2018}; (3) \citet{Blake2010}; (4) \citet{Suarez2023}.
}
\end{table*}

\noindent Table~\ref{table:variability} list parameters of interest (including rotational periods, variability amplitudes and inclination) of our sample. The following paragraphs elaborate on parameters for each object in our sample.

\textit{\obj{0047}} \citep{Gizis2012,Gizis2015} is an unusually red \citep[$J-K_s = 2.5\pm0.1$ mag;][]{Best2021} dwarf with optical spectral type L7 $\gamma$ and IR type L6-8 $\gamma$ \citep{Gizis2012,Gizis2015}. It has one of the highest known variable amplitudes observed with both the Hubble Space Telescope (HST) \citep[11\% at 1.1 $\mu$m;][]{Lew2016} and the \textit{Spitzer} Space Telescope ($1.07\pm 0.04$\% at 3.6 $\mu$m) with a rotational period of $16.4\pm0.2$ h \citep{Vos2018}. It is viewed equator-on with a measured inclination of $i={85^{+5}_{-9}}^{\circ}$ \citep{Vos2018}.

\textit{\obj{0355}} \citep{Reid2008} is an extremely red \citep[$J-K_s = 2.438\pm0.004$;][]{Lawrence2007, Schneider2023} object with optical spectral type L5 $\gamma$ \citep{Cruz2009} and IR type L3-6 $\gamma$ \citep{Gagne2015}. Significant variability ($0.26\pm0.02$\%) with a period of $9.53\pm0.19$ h was measured in the 3.6 $\mu$m band with \textit{Spitzer} \citep[][]{Vos2022}. \citet{Suarez2023} measured an inclination angle of $i=50\pm 2 ^{\circ}$.

\textit{\obj{0642}} \citep{Mace2013} was discovered as an "extremely red" object in the H band and later classified as an IR L9 \citep{Best2015}. It is not as red in the NIR with $J-K_s = 1.86\pm0.02$ mag \citep{Schneider2023}. It has the highest measured variability amplitude at 3.6 $\mu$m of $2.16\pm0.16$\% with a period of $10.11\pm0.06$ h from \textit{Spitzer} \citep[][]{Vos2022}.

\textit{\obj{1741}} \citep{Schneider2014} is an IR L6-8 $\gamma$ \citep{Gagne2015,Faherty2016} dwarf with a particularly red NIR colour of $J-K_s = 2.56\pm0.05$ mag \citep{McMahon2013,Best2021}. It has detected variability with a low amplitude ($0.35\pm0.03$\% at 3.6 $\mu$m) with a period of $15.00^{+0.71}_{-0.57}$ h \citep[][]{Vos2022}.

\textit{\obj{2206}} \citep{Kirkpatrick2000} is an optical and IR L4 $\gamma$ \citep{Gagne2015,Faherty2016} dwarf. It is not particularly redder than other L dwarfs with similar magnitudes with $J-K_s = 1.9\pm0.1$ mag \citep{Best2021}. It has a non-detection of variability from a \textit{Spitzer} light curve at 3.6 $\mu$m \citep{Vos2022}.

\textit{\obj{2244}} \citep{Dahn2002} is a very red \citep[$J-K_s = 2.57\pm0.02$ mag][]{Schneider2023} optical L6.5 \citep{AllersLiu2013} and IR L6-8 $\gamma$ \citep{Faherty2016} object that is spectroscopically similar to \obj{0047} \citep{Gizis2015}. Variability has been detected by \textit{Spitzer} at 4.5 $\mu$m \citep{MoralesCalderon2006} and more recently, at 3.6 $\mu$m with an amplitude of $0.8\pm0.2$\% and rotational period of $11\pm2$ h \citep{Vos2018}. There has also been variability detected in the J-band light curve with an amplitude of 5.5\% \citep{Vos2019}. This object has an inclination angle of $i={76^{+14}_{-20}}^{\circ}$ \citep{Vos2018}.

\section{Observations and data reduction}
\label{sec:observations}
\subsection{JWST}
\label{sec:JWSTobservations}
We observed our six objects using JWST MIRI \citep[][]{Rieke2015} and Near Infrared Spectrograph \citep[NIRSpec;][]{Jakobsen2022} instruments to obtain near-simultaneous $R\sim100$ spectra from 0.6 -- 14 $\mu$m (PI: Vos, GO 3486). The same detector setup was used for all targets. Only the exposure times varied according to the brightness of each target. A summary of the observations can be found in Table \ref{table:observations}.

\begin{table*}
\caption{Summary of JWST observations.}             
\label{table:observations}      
\centering          
\begin{tabular}{c c llll }    
\hline\hline       
\noalign{\smallskip}

Obs. Start Time (UT) & Target & Instrument & Groups in & Exposure Time (s) & S/N\\
&&&integration&& MIRI at 6 $\mu$m\\
&&&&&NIRSPEC at 2 $\mu$m\\
\hline
\noalign{\smallskip}

2024-02-09 04:46:17 & \obj{0047} & MIRI & 13 & ~~36.076 & 491\\
2024-02-09 05:12:55 & \obj{0047} & NIRSPEC & 4 & ~~~~6.232 & 269\\
2024-02-17 14:06:20 & \obj{0355} & MIRI & 6 & ~~16.650 & 451 \\
2024-02-17 14:36:24 & \obj{0355} & NIRSPEC & 2 & ~~~~3.116 & 324\\
2024-03-07 12:44:09 & \obj{0642} & MIRI & 35 & ~~97.126 & 446\\
2024-03-07 13:12:35 & \obj{0642} & NIRSPEC & 6 & ~~~~9.348 & 187\\
2024-04-05 03:27:26 & \obj{1741} & MIRI & 25 & ~~69.376 & 553\\
2024-04-05 03:55:54 & \obj{1741} & NIRSPEC & 5 & ~~~~7.790 & 249\\
2024-06-09 22:54:50 & \obj{2206} & MIRI & 45 & 124.877 & 525\\
2024-06-09 23:34:11 & \obj{2206} & NIRSPEC & 8 & ~~12.464& 332 \\
2024-06-21 16:19:37 & \obj{2244} & MIRI & 25 & ~~69.376 & 484\\
2024-06-21 16:47:13 & \obj{2244} & NIRSPEC & 8 & ~~12.464 & 233\\
\hline                  
\end{tabular}
\end{table*}

We used the MIRI Low Resolution Spectrometer (LRS) fixed-slit mode (4.7" $\times$ 0.51") to obtain 5 -- 14 $\mu$m spectra with a resolving power varying from $R \sim 40 - 160$ for each target. We used the default P570L disperser and FULL subarray with the FAST readout pattern. We also used the F1000W filter during target acquisition. We used the two-point along slit AB nod pattern to obtain two total dithers. The exposure time for each target (see Table~\ref{table:observations}) was calculated to ensure a S/N > 75 at 10 $\mu$m for all targets, so that the silicate feature at 8 -- 11 $\mu$m can be robustly analysed.

During the same visit as our MIRI observations, we obtained NIRSpec spectra between 0.6 -- 5.3 $\mu$m for the same target. NIRSpec was used in fixed-slit mode with the Prism/CLEAR grating/filter pair, the S200A1 slit (0.2" $\times$ 3.2") and the NRSRAPID readout pattern. This filter provided a spectral resolution of $R\sim100$. We observed in an AB nod pattern with four total dithers. With the exposure times in Table~\ref{table:observations}, we achieved S/N > 100 across the 1-5.2 $\mu$m wavelength range, which is sufficient to constrain atmospheric properties from the water, methane and carbon monoxide features.

\subsubsection{Data reduction}
\label{sec:data-reduction}
We used the JWST calibration pipeline version 1.16.0 \citep{Bushouse2024} with CRDS context file \verb|jwst_1322.pmap| to reduce the JWST data, starting from uncalibrated data downloaded from the Mikulski Archive for Space Telescopes (MAST). Default parameters were used when processing the spectra through each of the three stages.\\
\\
Stage 1 of the pipeline applies detector-level corrections and ramp fitting to the uncalibrated ramp data for each exposure. This includes dark and bias subtraction, removal of bad pixels and cosmic ray detection, for example. This stage produces two-dimensional count rate images for each exposure. Stage 2 uses the uncalibrated slope images produced by stage 1 to perform instrument calibrations for each exposure. This includes pixel flat-fielding, flux calibration and assigning world coordinate information. This stage outputs calibrated slope images. Finally, Stage 3 combines each of the corrected dithers into a single science-ready spectrum. We use the final \verb|x1d.fits| file in our data analysis throughout this paper. This was repeated for each object for both NIRSpec Prism and MIRI LRS to obtain a full 0.6 -- 14 $\mu$m spectrum.

\subsubsection{Flux calibration and merging of spectra}
To create a continuous 0.6--14 $\mu$m spectrum for each object, the NIRSpec Prism and MIRI LRS spectra were merged. At the overlap window of 5 -- 5.3 $\mu$m, we monitored the S/N of the NIRSpec spectra. When the S/N first drops below a threshold of 80 within the 5 -- 5.3 $\mu$m overlap region, we cut off the NIRSpec spectrum and switch to MIRI. This threshold was chosen to balance a high S/N with maintaining the higher spectral resolution provided by NIRSpec to resolve molecular features in this region, which would not be possible at this wavelength with MIRI LRS.

All objects except \obj{2244} have consistent flux in the NIRSpec and MIRI overlap region, which is expected due to the near-simultaneous observations. However, \obj{2244} has a flux offset between the NIRSpec and MIRI observations, where the NIRSpec spectrum has approximately 30\% less flux than the MIRI spectrum within the overlap region. The NIRSpec exposure started 27.6~min after the start of the MIRI exposure. This large difference cannot be explained by the object's variability \citep[0.8\% at 3.6~$\mu$m with $P=11.0\pm2.0$~h;][]{Vos2018} or any other physical atmospheric processes. We also compared synthetic photometry calculated from our JWST spectra with literature photometric points obtained from MKO \citep{Schneider2023} and WISE \citep{Lang2016}, which indicated that the NIRSpec spectrum had lower flux than expected. The flux from our MIRI spectrum matched well with literature photometry as well as archival \textit{Spitzer} IRS spectra \citep{SuarezMetchev2022}. Thus, we determined that the offset was possibly due to a slit misalignment from imperfect NIRSpec target acquisition. Objects in our sample have high proper motions (see Table~\ref{table:properties}), which may have contributed to the difficulty of acquiring the target. The NIRSpec target acquisition field of view is smaller than that of MIRI (see Sec.~\ref{sec:JWSTobservations}), which may have led to the difficulty in acquisition. To circumvent this issue, we normalised the NIRSpec spectrum to the flux in the overlapping region with MIRI LRS to create a continuous, high S/N 0.6 -- 14 $\mu$m spectrum.

\begin{figure*}
    \centering
    \includegraphics[width=\linewidth]{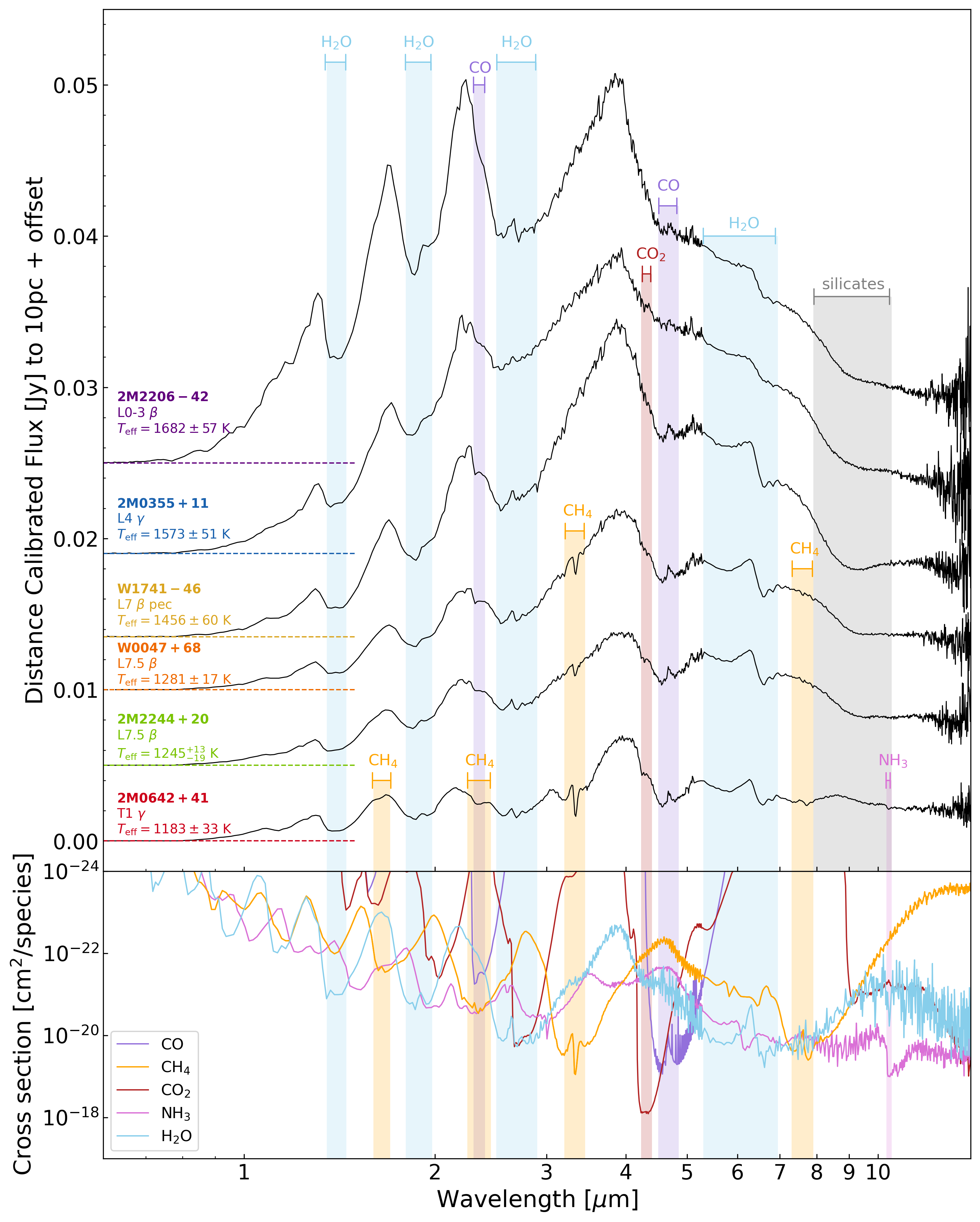}
    \caption{Distance-calibrated spectra for each object (top panel). Shaded regions show key absorption bands for \protect\obj{ch4}, CO, \protect\obj{co2}, \protect\obj{nh3} and \protect\obj{h2o}. The dotted line shows the offset level for each spectrum. The spectra are ordered by decreasing effective temperature from top to bottom. The cross sections for each species, computed at $T = 1300$ K and $P = 1$ bar, are included in the bottom panels.}
    \label{fig:opacities}
\end{figure*}

After data reduction, we used \verb|SEDkit| \citep{Filippazzo2015} to normalise each spectra to 10 pc using the measured parallaxes from the literature (shown in Table~\ref{table:properties}). We present the combined NIRspec and MIRI spectra for all objects in our sample in Fig.~\ref{fig:opacities}.

\subsection{High resolution spectra}
IGRINS is a high resolution ($R\sim 45\,000$) NIR spectrograph on Gemini South \citep{Yuk2010,2014Park,Mace2016,Mace2018}. IGRINS has two separate spectrograph arms to cover the H and K bands, resulting in full coverage of the wavelength range between 1.45 -- 2.45 $\mu$m.

We obtained IGRINS spectra for the two objects with missing inclination measurements (PI: Vos, PID: GS-2020B-Q-246). \obj{1741} was observed on three nights at UTC 2021-01-24 08:57:06, 2021-01-27 08:53:52 and 2021-01-28 08:57:37. All observations had an integration time of 350 s. The observation on 2021-01-24 had low S/N $\sim 10$ at and the observation on 2021-01-27 had S/N $\sim 30$, likely due to high air mass. The observation on 2021-01-28 had a S/N $\sim 300$ in the centre of the order of interest (77), so we used this night's spectrum for our final results and analysis. \obj{2206} was observed on UTC 2020-11-16 01:46:26 for an integration time of 200 s, resulting in a max S/N $\sim 30$ in the middle of order 77.

Reduced IGRINS spectra were downloaded using version 3 of the Raw and Reduced IGRINS Spectral Archive (RRISA) \citep{kaplan2024,Sawczynec2025}. The IGRINS Pipeline Package \citep{kaplan2024} removes effects due to cosmic rays, corrects for instrumental flexures, stacks individual exposures, removes sky emission features and the detector readout pattern, applies flat fielding, corrects individual echelle orders, calculates the wavelength solution, and extracts the flux per pixel and variance spectra.\\
\\
A high-resolution GNIRS (Gemini Near-InfraRed Spectrograph) spectrum for \obj{0642} is also publicly available (PI: Best, PID: GN-2015B-FT-14). This spectrum was observed on UTC 2016-1-09. However, when analysing this spectrum, we noticed that the spectrum had a very low S/N, so we were unable to measure an accurate $v\sin i$ or RV. With better quality spectra of \obj{0642}, we would be able to both confirm its moving group membership, as well as measure its $v\sin i$ and inclination angle.

\section{Spectral, fundamental and kinematic parameters}
\subsection{Updated spectral typing}
\label{sec:spectral-typing}
We reclassified the spectral types of the objects in our sample by comparing the recently obtained NIRSpec Prism spectra to field gravity ($\alpha$), intermediate gravity ($\beta$) and very low gravity ($\gamma$) NIR spectral standards for L \& T dwarfs (\textcolor{cobalt}{Bickle et al. in prep.}). As there are variations in the NIR slope between our sample and the spectral standards, each object's \textit{J}-, \textit{H}- and \textit{K}-bands were individually dereddened using the \cite{Cardelli1989} extinction law. For each band, $E(B-V)$ (colour excess due to reddening) and $R_V$ (total-to-selective extinction ratio) values were adjusted until the best $\chi^2$ fit was achieved to the respective band of each standard, and the resulting fits were visually compared to determine the best spectral type match.

{The benefit of this dereddening approach vs. simple band-by-band normalisation (e.g., \citealt{Cruz2018}) is that some objects can appear redder even within a spectral band ($J$, $H$ or $K$), not just across bands, due to peculiar composition, clouds or youth. This is especially true near the L/T transition where field objects can show a wide range of slopes even within a single band, within one spectral subtype. This scheme therefore allows us to place objects more robustly on the grid of spectral standards and assess for peculiarities independent of slope changes. Interstellar extinction laws have previously been shown to improve the spectral fits of young brown dwarfs to model spectra \citep{Petrus2020,Petrus2024,Hurt2024,Mader2026}, suggesting that they may indeed act as appropriate analogues for the extinction within young substellar atmospheres. Nonetheless, we treat the extinction parameters providing the best fits to our objects as nuisance parameters and not as physically relevant quantities.}

\begin{figure}
    \centering
    \includegraphics[width=\linewidth]{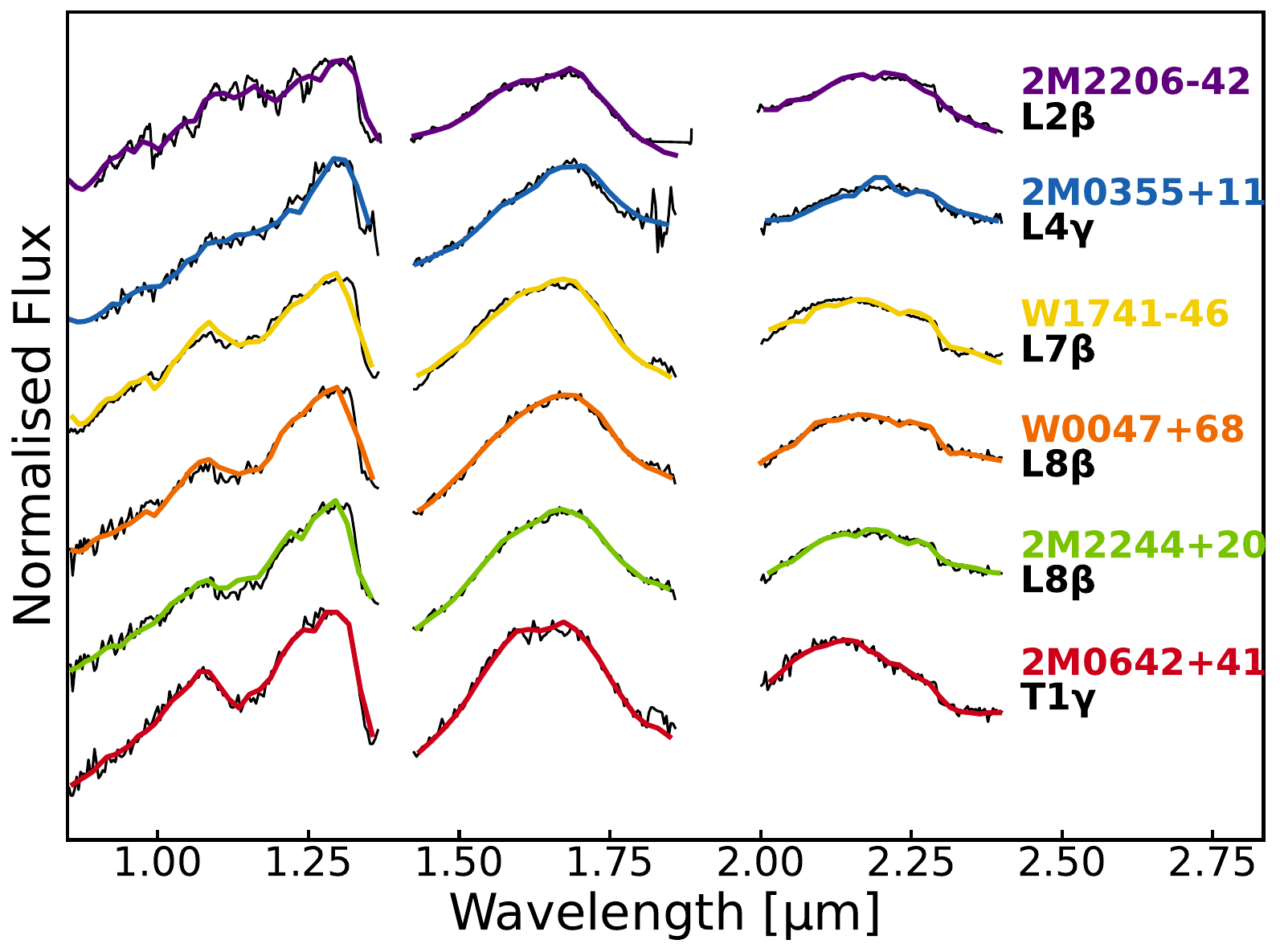}
    \caption{{JWST/NIRSpec Prism spectra of our sample (coloured) compared to their respective best fitting spectral standard (black) after dereddening (\textcolor{cobalt}{Bickle et al. in prep.}).}}
    \label{fig:sptfits}
\end{figure}

{The dereddened NIRSpec Prism spectra for our sample are plotted against their respective best fitting spectral standard in Figure~\ref{fig:sptfits}} and we present our updated spectral types in Table~\ref{table:fundametal-params}. {The adopted spectral types in Table~\ref{table:fundametal-params} do not necessarily match the best fitting spectral standards presented in Fig.~\ref{fig:sptfits}, as the fits of some objects are in between multiple spectral standards, which results in a range of spectral types in Table~\ref{table:fundametal-params}.} 

\begin{figure}
    \centering
    \includegraphics[width=\linewidth]{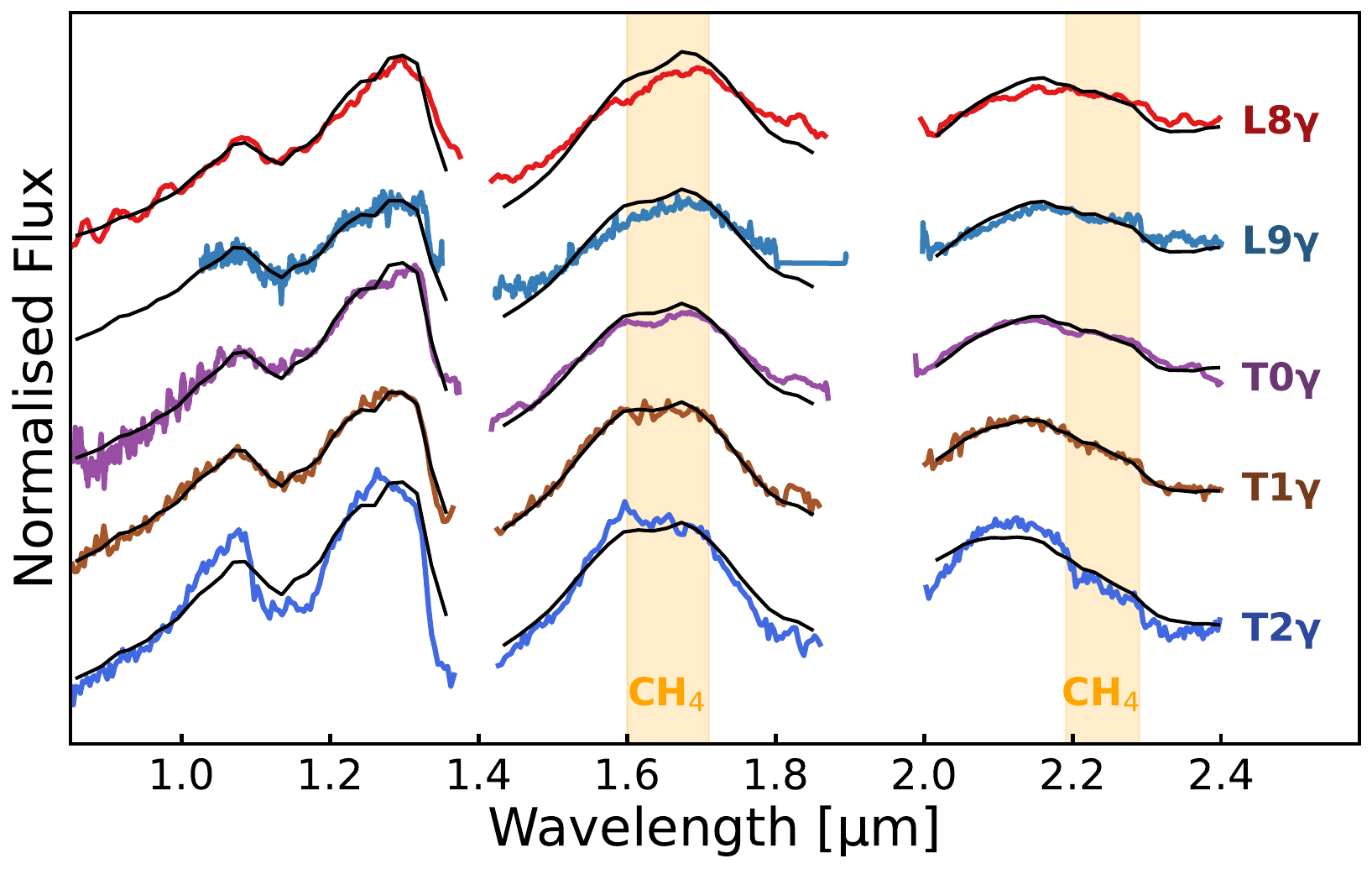}
    \caption{{JWST/NIRSpec Prism NIR spectrum of \protect\obj{0642} (black), dereddened and compared to L/T transition low-gravity spectral standards. The T1$\gamma$ spectral standard is the best fit to \protect\obj{0642}. CH$_4$ absorption regions are shaded in orange.}}
    \label{fig:0642spt}
\end{figure}

Of significant note is \obj{0642}, which was previously classified as L9 \citep{Best2015}, and is now confirmed to instead be a T1 object. {Figure~\ref{fig:0642spt} shows its NIRSpec spectrum dereddened and compared to L8-T2$\gamma$ spectral standards. Note that the overall continuum shape and, crucially, the CH$_4$ absorption bands which define the transition into the T dwarfs \citep{Burgasser2002}, are best fit to the T1$\gamma$ standard.} {Our updated spectral classification of \obj{0642} as T1 is supported by its bluer position on the colour-magnitude diagram (see Fig.~\ref{fig:colourmag}). }

The absence of the mid-IR silicate absorption feature in \obj{0642} is consistent with its updated spectral classification. At the L/T transition, silicate clouds sink below the photosphere \citep{AckermanMarley2001,SaumonMarley2008,Cushing2006,SuarezMetchev2021}.

\subsection{Fundamental parameters}
\label{sec:fundamental-parameters}
SEDs can be used to determine bolometric luminosities ($L_{\text{bol}}$) and combined with age estimates and evolutionary models, we can estimate the fundamental parameters (i.e. effective temperature ($T_{\text{eff}}$), radius, mass ($M$) and surface gravity ($\log g$)) of brown dwarfs \citep[e.g.][]{Filippazzo2015,Best2021,Kirkpatrick2021}. These parameters further characterise our sample and place them in context with a wider sample of brown dwarfs. {Physical parameters derived from the bolometric luminosity and ages from SEDs depend on evolutionary models, rather than atmospheric models, which can have systematic uncertainties introduced from their sensitivity to boundary conditions} \citep{Filippazzo2015,Suarez2021,Sanghi2023}.

We constructed entire SEDs for each object in our sample by combining our 0.6--14 $\mu$m spectra and existing  photometric points from 2MASS and WISE. We used \verb|SEDkit| \citep{Filippazzo2015} to interpolate the \citet{SaumonMarley2008} evolutionary model to the SEDs to obtain their fundamental parameters. The $L_{\text{bol}}$ for each object was calculated by integrating under the flux-calibrated SEDs. \verb|SEDkit| then uses the $L_{\text{bol}}$ and the object's age ($133^{+15}_{-20}$ Myr for AB Dor members; \citealt{Gagne2018}) to estimate the radius and $M$ from the \citet{SaumonMarley2008} evolutionary model isochrones. {Since \obj{0642} is likely a member of the Oceanus moving group (see Section \ref{sec:sample-properties}), we use the age of this moving group \citep[$510\pm95$ Myr;][]{Gagne2023} to calculate the fundamental parameters instead.} $T_{\text{eff}}$ was calculated by combining the radius and $L_{\text{bol}}$ with the Stefan-Boltzmann law. Our measured values for each object are presented in Table~\ref{table:fundametal-params}. 

\begin{table*}
    \caption{Fundamental parameters from SED analysis with updated spectral types for each object}
    \label{table:fundametal-params}
\centering          
\begin{tabular}{lcccccc }     % 7 columns 
\hline\hline       
\noalign{\smallskip}
Object & NIR Spectral type &$L_{\text{bol}}$ [$\log (L_{\text{bol}}/L_{\odot} )$] & $T_{\text{eff}}$ [K] & Radius [R$_{\text{Jup}}$] & $M$ [M$_{\text{Jup}}$] & $\log g$\\
\hline
\noalign{\smallskip}

\obj{0047} & L7.5 $\beta$ & $-4.417\pm0.018$ & $1281\pm17$ & $1.22\pm0.03$ & $17.2\pm2.1$ & $4.46^{+0.07}_{-0.08}$ \\
\obj{0355} & L4 $\gamma$ &$-4.072\pm0.006$ & $1573\pm51$ & $1.21\pm0.08$ & $30.3\pm6.8$ & $4.71\pm0.17$ \\
\obj{0642} & T1 $\gamma$&$-4.687\pm0.044$ & {$1183\pm33$} & {$1.045\pm0.03$} & {$26.7\pm2.8$} & {$4.78\pm0.06$} \\
\obj{1741} & L7 $\beta$ pec&$-4.201\pm0.048$ & $1456\pm60$ & $1.21\pm0.08$ & $27.2\pm6.8$ & $4.65\pm0.18$ \\
\obj{2206} & L0-3 $\beta$&$-3.957\pm0.033$ & $1682\pm57$ & $1.20\pm0.07$ & $33.8\pm6.8$ & $4.75\pm0.16$ \\
\obj{2244} & L7.5 $\beta$&$-4.478\pm0.015$ & $1245^{+13}_{-19}$ & $1.21\pm0.02$ & $16.1\pm1.5$ & $4.44^{+0.05}_{-0.06}$ \\
\noalign{\smallskip}
\hline
\end{tabular}
\tablefoot{$T_{\text{eff}}$, radius, $M$ and $\log g$ for \obj{0047} and \obj{2244} were obtained from additional rejection sampling (see Appendix~\ref{appendix:hist}). {The fundamental parameters for \obj{0642} were calculated using the age of the Oceanus moving group \citep[$510\pm95$ Myr;][]{Gagne2023}.}}
\end{table*}

We have high quality JWST spectra covering the wavelength range between 0.6--14 $\mu$m, so we can construct the best possible SEDs for these objects, and thus calculate the most precise measurements for the fundamental parameters. We calculated our SEDs to be $> 98$\% complete by comparing our SED to Sonora Diamondback models \citep{Morley2024} using the \verb|SEDA| (Spectral Energy Distribution Analyzer) package \citep[][ \textcolor{cobalt}{Suárez et al. in prep.}]{Suarez2021}. Our measured $T_{\text{eff}}$, radii and $M$ (see Table~\ref{table:fundametal-params}) are consistent with previously measured values from \citet{Filippazzo2015,Vos2022}. 

At $\sim133$ Myr, \obj{0047} and \obj{2244} are close to the deuterium burning limit of $\sim12-13 \text{ M}_{\text{Jup}}$, meaning that we expect a bimodal distribution of masses, similar to VHS~1256~b \citep{Miles2023,Dupuy2023}. We implemented a similar rejection sampling method to \citet{Miles2023} with our $L_{\text{bol}}$ to obtain a more robust measurement of the fundamental parameters for \obj{0047} and \obj{2244}. We include the histograms for our posteriors in Appendix \ref{appendix:hist}.\\
\\
In Fig.~\ref{fig:fundamental-params}, we compare the $T_\text{eff}$ and $M$ of our sample to a larger sample (> 1000) of ultracool dwarfs, including older field dwarfs from \citet{Sanghi2023}. \citet{Sanghi2023} measured their $L_{\text{bol}}$ by integrating optical to mid-IR SEDs and used the \citet{Baraffe2015} and \citet{SaumonMarley2008} evolutionary models to determine the fundamental parameters for their sample. We notice that there is a large spread in our sample of brown dwarfs. Objects with a later spectral type have a cooler $T_{\text{eff}}$, which agrees with the general trend observed from \citet{Sanghi2023}. More massive brown dwarfs are hotter than those of the same spectral type. Younger objects also tend to have lower $T_{\text{eff}}$ compared to older field objects of the same spectral type \citep{Faherty2016,Sanghi2023} (see polynomial relation for young objects in Fig.~\ref{fig:fundamental-params}). Most of our objects tend to follow the trend for young objects, apart from \obj{1741}, which appears to be closer to the old field relation, despite being a bona fide AB Dor member.

\begin{figure}
    \centering
    \includegraphics[width=\linewidth]{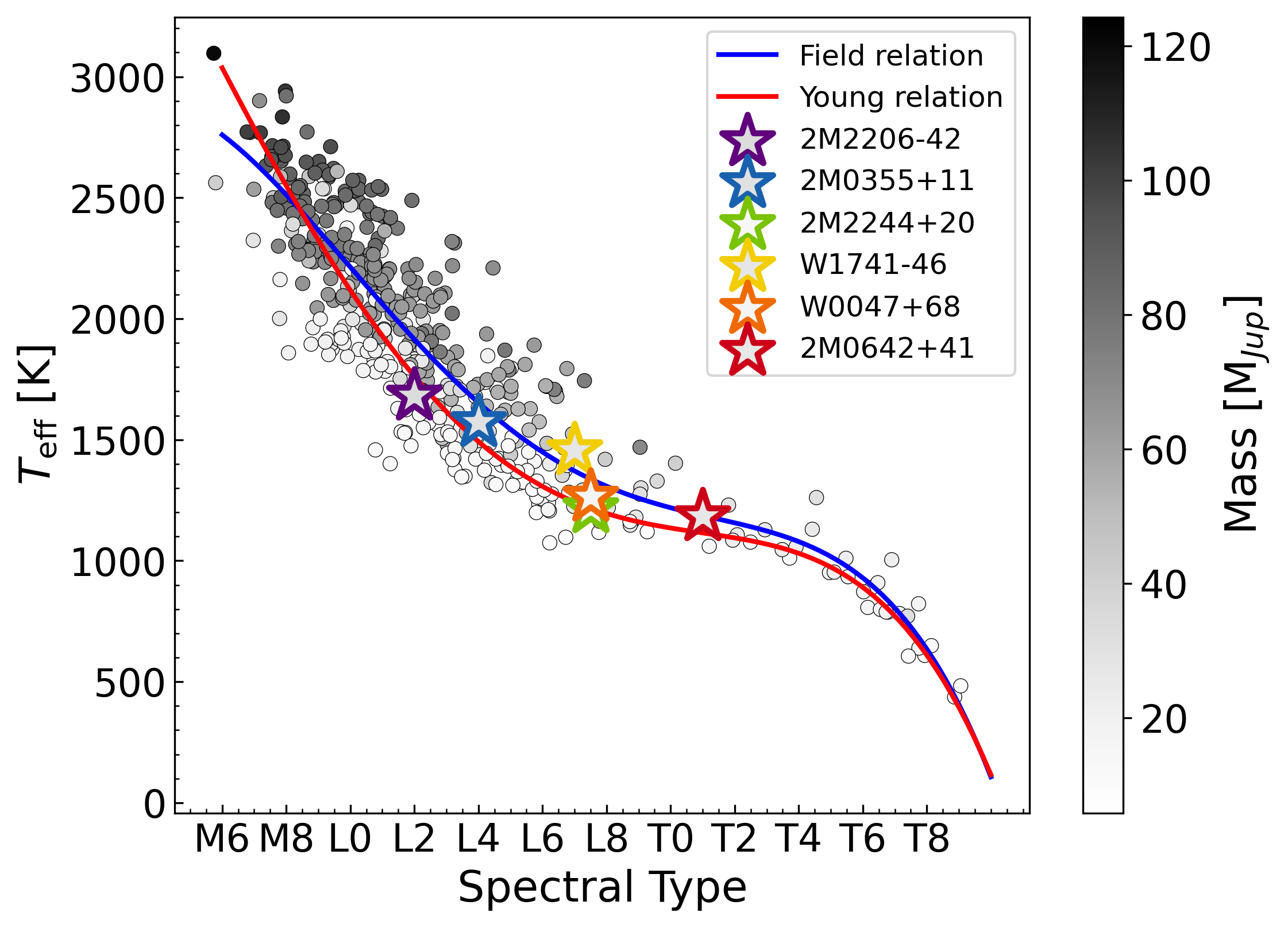}
    \caption{Visualisation of the fundamental parameters of the sample analysed in \citet{Sanghi2023} (grey circles) and our sample (coloured outlined stars). $T_{\text{eff}}$ is plotted against IR spectral type. The fill colour is indicative of the mass of the object. Random noise of 0.3 spectral types was added along the x-axis to minimise overlapping points in visualisation. The fundamental parameters were calculated using evolutionary models. The polynomial relations for the old field dwarfs and young moving group members (blue and red lines, respectively) from \citet{Sanghi2023} are overplotted.}
    \label{fig:fundamental-params}
\end{figure}

\subsection{Rotational velocities}
\label{sec:igrins}
\begin{figure*}
\centering
\includegraphics[width=\textwidth]{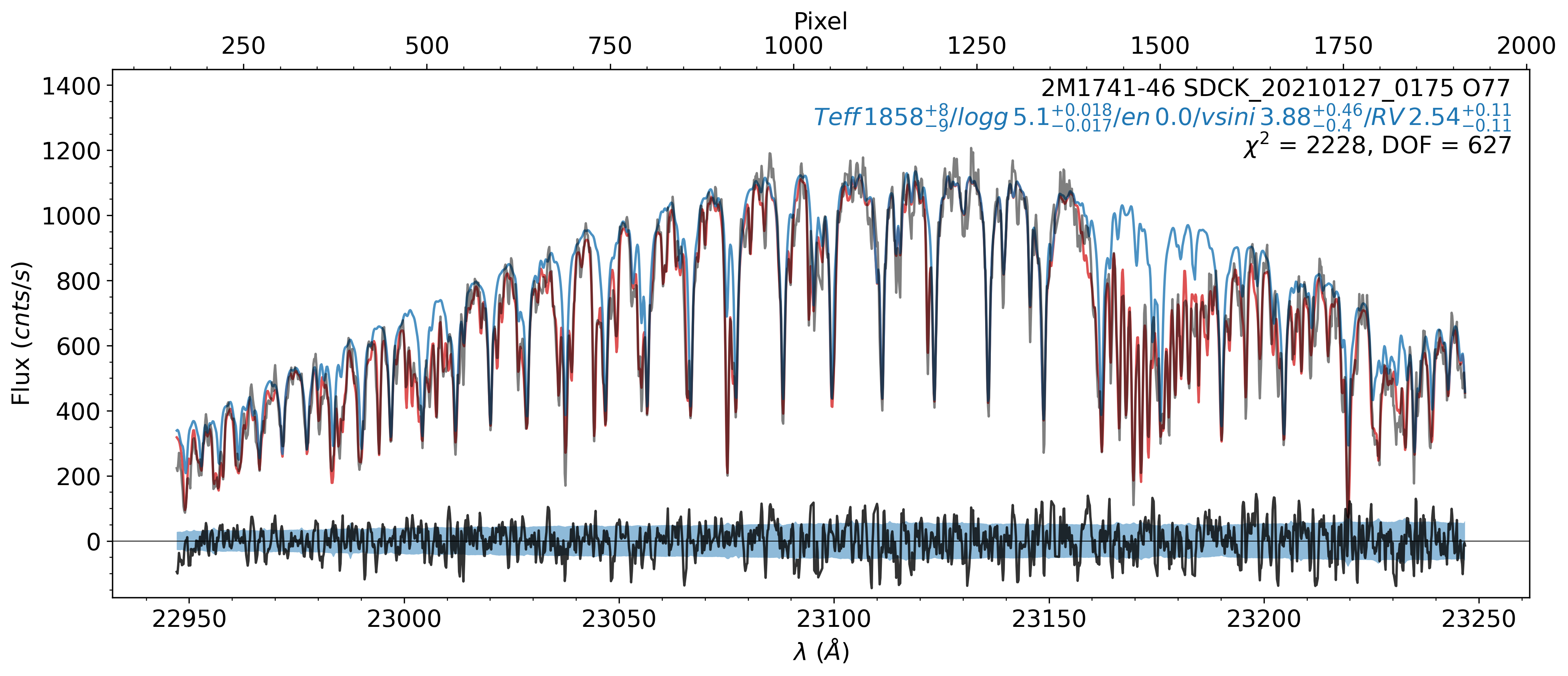}
\includegraphics[width=\textwidth]{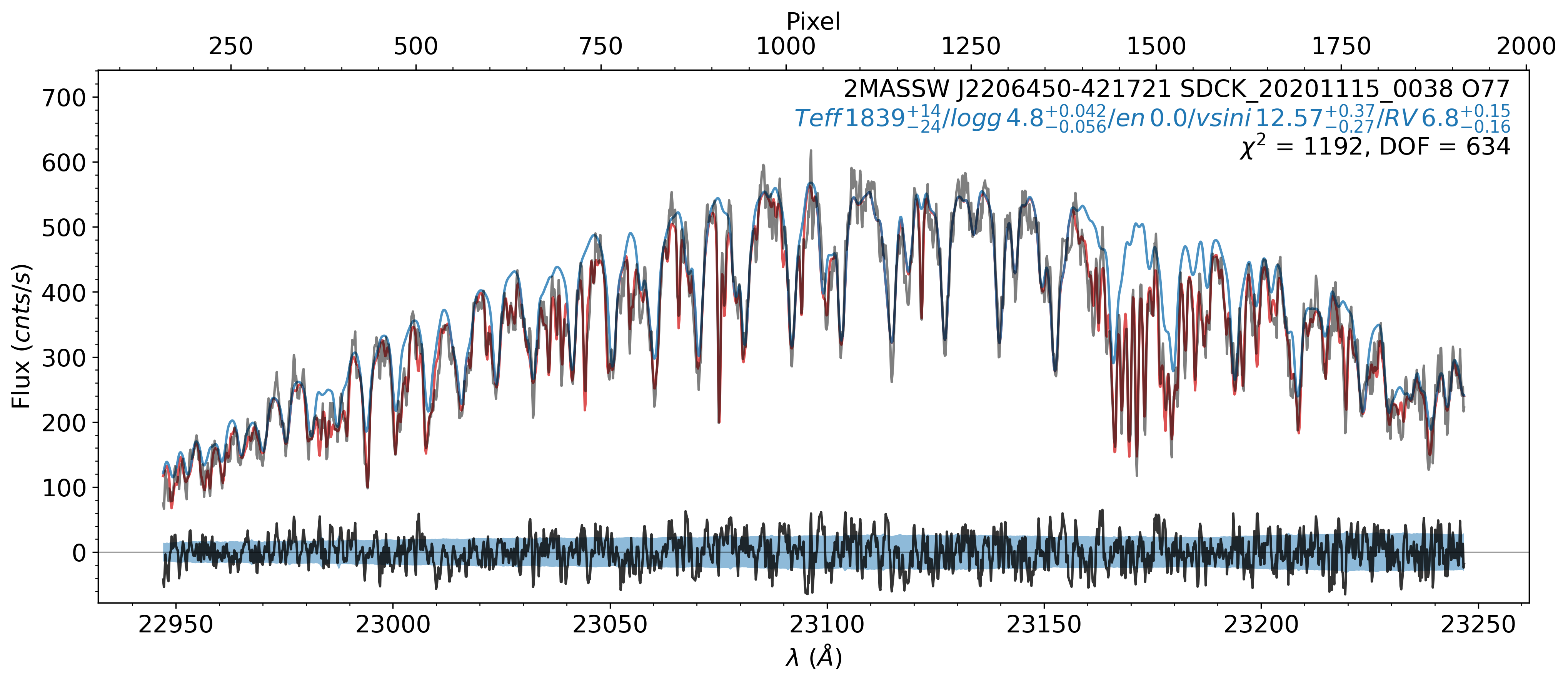}
\caption{SMART fits to IGRINS spectra for \protect\obj{1741} (top) and \protect\obj{2206} (bottom). The grey line is the observed IGRINS spectra, the red line is the fitted model including telluric absorption, and the blue line is the fitted model without tellurics. At the bottom of each plot is the residual between the model (including tellurics) and the observed data. The shaded blue region is the $\pm 1\sigma$ uncertainty.}
\label{fig:igrins}
\end{figure*}

Previous studies have shown that there is a strong latitudinal dependence of clouds and infrared colours, where objects viewed equator-on appear redder than those viewed pole-on \citep{Vos2017,Vos2020,SuarezMetchev2022,Suarez2023}. This colour variation is believed to be caused by a larger vertical extent of clouds along the equator compared to the poles \citep{TanShowman2021}. The larger vertical extent would lead to higher clouds at the equator, which are expected to be cooler and composed of smaller particles \citep{AckermanMarley2001}. We aim to test this latitudinal dependence in our sample of young, coeval AB Dor members, removing any potential biases due to age. We have previous measurements for variability and inclination for four targets (\obj{0047}, \obj{0355}, \obj{0642}, and \obj{2244}) (see Table~\ref{table:variability}). We used our high resolution IGRINS spectra for the remaining two targets \obj{1741} and \obj{2206} to measure their $v\sin i$, which is needed to determine their inclinations.

We used the Spectral Modelling Analysis and RV Tool (SMART) \citep{Hsu2021} to measure the $v\sin i$ for our remaining targets. SMART is a Markov Chain Monte Carlo (MCMC) tool that can be used for high-resolution near-infrared spectra, including IGRINS. SMART uses a forward modelling method to simultaneously fit the wavelength solution, telluric line depths and the Full Width Half Maximum (FWHM) of the instrumental Line Spread Function (LSF) of the standard star, and then uses this best-fit model with atmospheric models on the observed spectra to fit the free parameters for our target which include $T_{\text{eff}}$, $\log g$, $v\sin i$ and RV. 

We ran SMART on order 77 of the K-band spectrum. This order contains many CO absorption lines located in the wavelength range 2.293 -- 2.325 $\mu$m. We chose to use the PHOENIX BT-Settl model atmosphere \citep{Allard2012} to fit our spectra. We used 50 walkers and ran the MCMC for 600 burn-in steps. For \obj{2206}, we ran the MCMC for 600 steps, and for \obj{1741}, we increased it to 1200 to ensure that the walkers had converged by visually inspecting the trace and ensuring the output results were not changing significantly. We fit the L dwarf parameters $T_{\text{eff}}$, $\log g$, $v\sin i$ and RV, as well as the parameters for instrumental or environmental effects: air mass (AM), precipitable water vapour (pwv), flux offset ($C_{F_\lambda}$), wavelength offset ($C_\lambda$) and the noise factor ($C_{\text{noise}}$). We set our prior ranges based on typical values for our L dwarf sample, with $T_{\text{eff}} =  1000 - 2000$ K, $\log g = 4-5$, $v\sin i = 0-50$ km/s and RV = $-50-50$ km/s. We present our new $v\sin i$ measurements in Table~\ref{table:variability} and our final fits to the spectra in Fig.~\ref{fig:igrins}.\\
\\
{The fitted $T_{\text{eff}}$ and $\log g$ are less accurate than those determined from our earlier SED analysis in Section~\ref{sec:fundamental-parameters} as they are derived from atmospheric models with intrinsic systematic offsets and are unreliable for young L/T transition objects and tend to report values higher than SED measurements \citep{Liu2013,Allers2016, Hsu2021}. These physical parameters are also only determined from a single order, instead of from a full SED. The narrow wavelength range provided by the high resolution spectrum is not reliable for estimating $T_{\text{eff}}$ \citep{Vos2018}. Because of this, the small error bars are only due to internal errors in fitting the model spectrum, and are not representative of the entire IGRINS spectrum. Since we were only interested in using the high resolution spectra to fit for $v\sin i$ and RV, fitting a single order was sufficient, and repeating this procedure across multiple spectral orders to obtain a more precise measurement for $T_{\text{eff}}$ and $\log g$ was deemed unnecessary as it was previously measured by SED analysis.}

Using the assumption that the brown dwarfs rotate as a rigid sphere, we can use our $v\sin i$ measurements with their measured periods (excluding \obj{2206}) and radii estimates from our SED analysis (see Sec.~\ref{sec:fundamental-parameters}) to determine their inclination angle, $i$. This follows the same assumption made in \cite{Vos2017}.
We report our newly measured inclination measurements for \obj{1741} and \obj{2206} in Table~\ref{table:variability}. \\
\\
Rotational velocities have been measured for a large number of L dwarfs, with a typical range of $v\sin i = 4-90$ km/s and a median $v\sin i = 20$ km/s \citep{Vos2017,Hsu2021}. Our $v\sin i$ measurement for \obj{1741} ($3.88^{+0.46}_{-0.40}$ km/s) was unusually low compared to the rest of our sample, and other previous $v\sin i$ measurements for similar L dwarfs. Our $v\sin i$ measurements were consistent across all three observations in the program. To verify this unusually low measurement, we checked our measurement of the LSF from our fit to the reference standard spectrum for each night. Our LSF measurement of $2.7^{+0.03}_{-0.04}$ km/s was consistent across all three observations of \obj{1741}, as well as the published IGRINS spectral resolution \citep[$R\sim45000$;][]{Yuk2010,2014Park}. These consistent LSF measurements indicate that the $v\sin i$ measurements are not biased by instrumental effects. This LSF measurement sets the lower limit of our $v\sin i$ measurement to 2.7 km/s. This indicates that our measurements are reliable and hence, we can confirm that our $v\sin i$ measurement is effectively constrained and can be trusted. Such a low $v\sin i$ measurement might suggest that \obj{1741} is viewed close to pole-on. Another possibility is that \obj{1741} may be slow- or non-rotating and viewed equator-on. If \obj{1741} was actually viewed equator-on, our $v\sin i$ measurement would correspond to a rotational period of $\sim 38$ h, which is greater than the expected rotational period of AB Dor members \citep[9 -- 17 h][]{Vos2022}. Since \obj{1741} has been previously observed to be variable by \citet{Vos2022} with a period of 15 h, we can estimate the inclination angle of \obj{1741} to be $i = 23.6 ^{+3.4}_{-3.6}{}^{\circ}$. 

\obj{2206} did not show significant variability in the \textit{Spitzer} light curve in \citet{Vos2022}, so we do not have an estimate for its period. Hence, we cannot constrain the inclination based on its $v\sin i$ alone. Instead, we have calculated the upper and lower bounds from the range of observed rotational periods in AB Dor, which is 9 -- 17 h \citep{Vos2022}. With a measured $v\sin i$ of 12.49 km/s and a minimum period of 9 hr, we find the lower limit for the inclination angle to be $i\sim49\degr$. With the maximum rotational period of 17 hr, $\sin i > 1$, so instead we set the maximum inclination angle to be $i=90\degr$, which corresponds to equator-on.

\subsection{Radial velocities}
We also measured new barycentric corrected RVs using SMART for \obj{1741} and \obj{2206} and present them in Table~\ref{table:properties}. {We used SMART's \verb|barycorr| function which uses \verb|astropy| to calculate the barycentric corrections. RVs were corrected given the time of each observation and Earth coordinates of Gemini South (30$^{\circ}$14.5’S, 70$^{\circ}$44.8’W).} These are the first RVs measured using high-resolution spectra from IGRINS for these objects. We provide the first RV measurement for \obj{2206} and an updated measurement for \obj{1741}. We calculated the updated membership probabilities with BANYAN $\Sigma$ to confirm \obj{1741} as a bona fide AB Dor member with a higher membership probability of 99.9\%, compared to a previous probability of 99.4\% \citep{Vos2022}. We provide the first RV measurement for \obj{2206}, which increases its membership probability to 99.8\% (previously 99.3\%), and confirms it as a bona fide AB Dor member with its full set of kinematic information.

\section{Key molecular absorptions}
\label{sec:key-molecular-absorptions}
Mid-L dwarfs are characterised by many strong molecular absorption bands, with some atomic and molecular bands (eg. Na, K and \obj{h2o}) strengthening as they cool to later spectral types, whereas other hydride bands such as FeH  weaken with cooler spectral types as iron clouds form \citep{Kirkpatrick2005,Cushing2006,Yamamura2010}. With JWST, we are able to observe a wide range of wavelengths near-simultaneously, allowing us to detect and characterise a broad array of molecular species for each target. This wide wavelength range also allows us to probe effects of nonequilibrium chemistry and reveal the vertical structure of the atmosphere \citep{Cushing2006}.

To identify key molecular absorption bands in the spectra, we consider cross sections of diverse chemical species, including water (\obj{h2o}), methane (\obj{ch4}), carbon monoxide (CO), carbon dioxide (\obj{co2}) and ammonia (\obj{nh3}). These cross sections were calculated at a temperature of 1300 K and a pressure of 1 bar with PICASO \citep{Batalha2020}. We chose to compute our cross sections at 1300K, as it is the approximate midpoint $T_{\text{eff}}$ of our sample. The cross sections were resampled to the same spectral resolution as our JWST spectra for ease of comparison. We show the absorption regions for these molecules in Fig.~\ref{fig:opacities}. In all objects, we visually identified \obj{h2o} between 1.33 -- 1.46, 1.78 -- 1.99, 2.48 -- 2.91 and 5.25 -- 6.95 $\mu$m, and CO between {2.3 -- 2.4 and} 4.5 -- 4.85 $\mu$m. However, only the three coolest dwarfs (\obj{0047}, \obj{2244} and \obj{0642}) show clear signatures of \obj{ch4} between 3.2 -- 3.45 $\mu$m and 7.3 -- 7.9 $\mu$m, and \obj{co2} absorption at 4.23 -- 4.4 $\mu$m. This agrees with previous observations of the onset of \obj{ch4} absorption typically around the L7 spectral type, and strengthening towards later spectral types \citep{Cushing2006,SuarezMetchev2022}. {\obj{0642} has the strongest \obj{ch4} absorption in our sample, with additional absorption present in the H- and K-bands between 1.6 -- 1.7 and 2.25 -- 2.45 $\mu$m.} At longer wavelengths for \obj{0642}, ammonia (\obj{nh3}) absorption is also visible at 10.3 -- 10.5 $\mu$m. {The \obj{ch4} and \obj{nh3} absorption bands in \obj{0642}} additionally reinforces its T spectral type classification. 

\subsection{Water and methane}
\begin{table}
\caption{\protect\obj{h2o}, \protect\obj{ch4} and silicate spectral indices.}             
\label{table:spectral-indices}      
\centering                          
\begin{tabular}{l c c c}        
\hline\hline        
\noalign{\smallskip}
Object & \obj{h2o} & \obj{ch4} & Silicate \\   
\hline  
\noalign{\smallskip}
   \obj{0047} & $1.127\pm0.009$ & $0.581\pm0.006$ & $1.43\pm0.05$ \\      
   \obj{0355} & $1.005\pm0.007$ & $0.572\pm0.009$    & $1.43\pm0.08$ \\
   \obj{0642} & $1.23\pm0.02$ & $0.95\pm0.01$     & $0.92\pm0.06$ \\
   \obj{1741} & $1.078\pm0.005$ & $0.537\pm0.009$    & $1.7\pm0.1$ \\
   \obj{2206} & $1.025\pm0.006$ & $0.564\pm0.008$   & $1.41\pm0.06$ \\ 
   \obj{2244} & $1.13\pm0.01$& $0.580\pm0.007$ & $1.45\pm0.05$\\
\hline                                   
\end{tabular}
\end{table}
\label{sec:waterandmethane}
\obj{h2o} and \obj{ch4} absorption dominate the spectra of mid-to-late L dwarfs, with increasing strength at later spectral types \citep{Cushing2006}. This trend was observed in a large sample of various ages, and we were interested whether our sample of young L dwarfs fit within this trend.

\begin{figure}
    \centering
    \includegraphics[width=\linewidth]{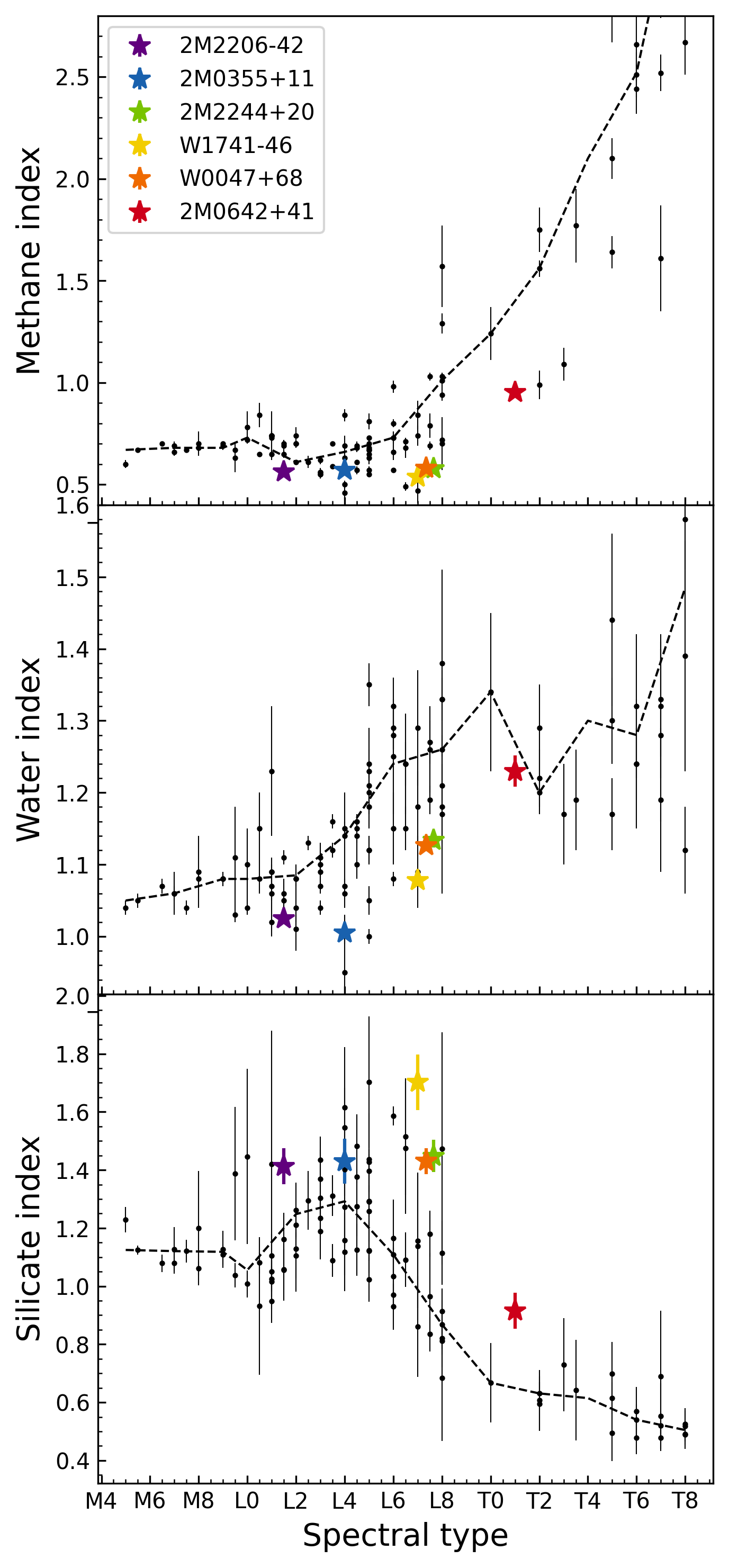}
    \caption{Calculated \protect\obj{ch4} (top), \protect\obj{h2o} (middle), and silicate (bottom) indices for our sample (coloured stars) compared to those from the \textit{Spitzer} IRS sample (shown as black points). The spectral types of \protect\obj{0047} and \protect\obj{2244} have an additional $-0.15$ and $+0.15$, respectively, for visualisation purposes. The dashed black line represents the median index for the bins of two spectral types in the \textit{Spitzer} IRS sample using our updated silicate index definition in the bottom panel.}
    \label{fig:indices}
\end{figure}

We measured the strength of the \obj{h2o} and \obj{ch4} absorption features at 6.25 $\mu$m and 7.65 $\mu$m using the \obj{h2o} and \obj{ch4} index defined in \citet{Cushing2006} and updated, for the case of \obj{ch4}, in \citet{SuarezMetchev2022}. The index is a ratio between the flux at a continuum level and the absorption feature. Thus, higher index values indicate stronger absorption features and vice versa. Using \verb|SEDA| \citep[][ \textcolor{cobalt}{Suárez et al. in prep.}]{Suarez2021}, we calculated the \obj{h2o} and \obj{ch4} indices for each target and compared to targets observed with \textit{Spitzer} IRS targets in \citet{SuarezMetchev2022} (see Fig. \ref{fig:indices}). We present our measured \obj{h2o} and \obj{ch4} indices in Table~\ref{table:spectral-indices}. Our later spectral type objects tend to have higher \obj{ch4} and \obj{h2o} absorption, following the general trend in \citet{Cushing2006}. For both \obj{h2o} and \obj{ch4}, the indices for our sample lie below the median index for the \textit{Spitzer} IRS sample, indicating that there is less absorption of \obj{h2o} and \obj{ch4} in our young objects compared to old field objects of similar spectral types.\\
\\
The shape of the water absorption features between 1 -- 4 $\mu$m also changes with temperature. The hotter three objects (\obj{2206}, \obj{0355} and \obj{1741}) show more triangular and "peaky" H, J and K bands, whereas the cooler three objects (\obj{0047}, \obj{2244} and \obj{0642}) have rounder features. The cooler objects have slightly lower surface gravities (see Table~\ref{table:fundametal-params}), which may suppress the shape of the H, J and K bands relative to the water opacities \citep{Faherty2016}. 

\subsection{Silicate absorption feature}
\label{sec:silicate-index}
\begin{figure*}
    \centering
    \includegraphics[width=\linewidth]{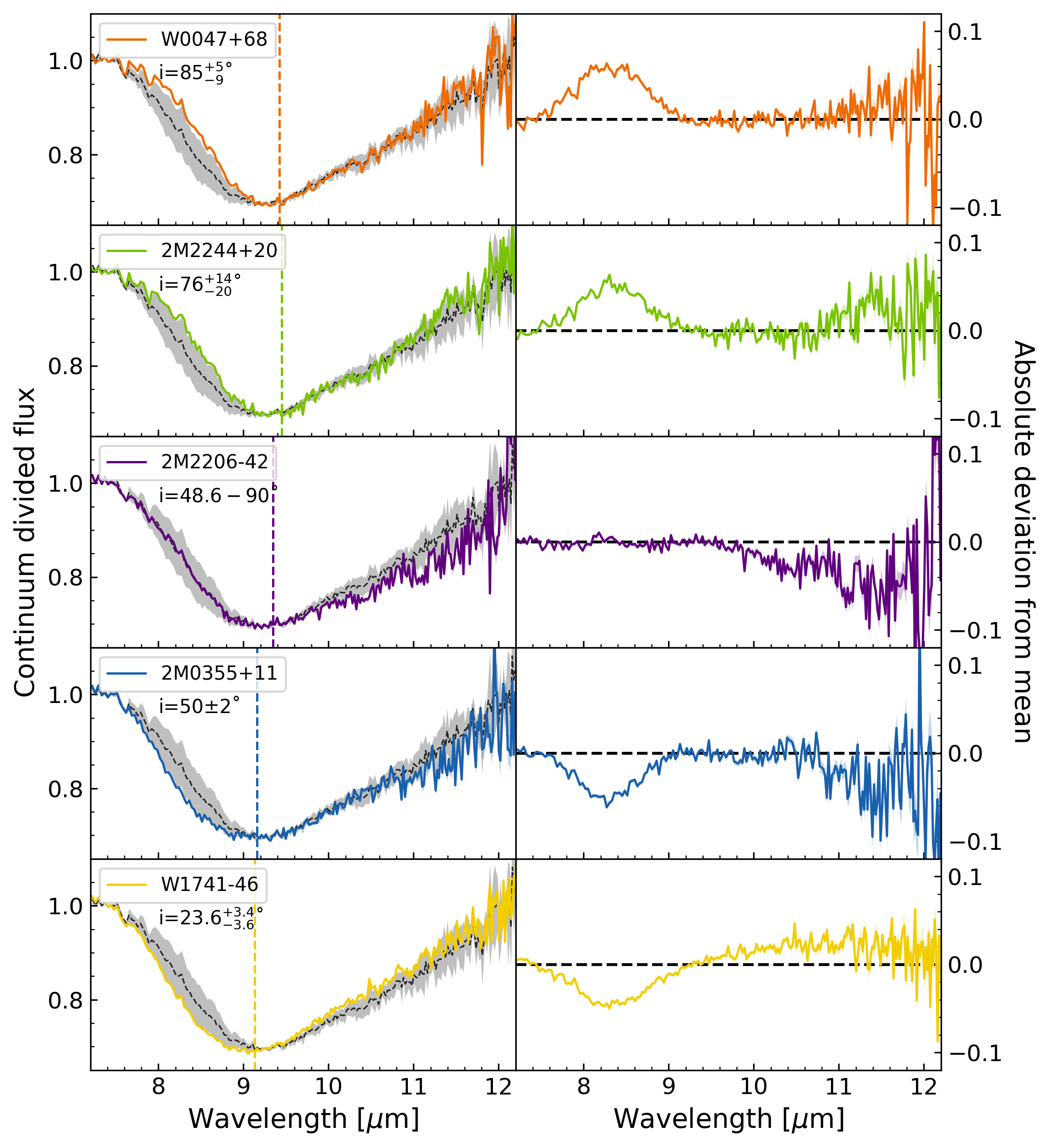}
    \caption{Isolated silicate absorption features of the objects in our AB Dor sample, after continuum removal. The spectra of the silicate feature has been ordered with inclination angle, with objects viewed more equator-on towards the top, and objects viewed more pole-on towards the bottom. The left panel directly compares the shape of the silicate feature with the mean silicate feature of our sample, shown as the black dotted line, with the shaded region representing one standard deviation. The vertical dashed line in the left panels shows the wavelength at peak absorption for each object. The right panels show the deviation from the mean silicate feature for each object.}
    \label{fig:silicate-feature}
\end{figure*}
The pressures and temperatures that exist in late-L dwarf atmospheres allow for the condensation of silicate grains in the visible photosphere \citep[e.g.][]{LoddersFegley2006,LunaMorley2021}. These condensates form silicate clouds that leave a prominent absorption feature that is detectable at 8 -- 11 $\mu$m \citep{LunaMorley2021}. All objects in our sample, except \obj{0642}, show clear signatures of this silicate absorption feature (see Fig.~\ref{fig:opacities}). Similar to the \obj{h2o} and \obj{ch4} index, the silicate index \citep{SuarezMetchev2022,SuarezMetchev2023} estimates the strength of the silicate absorption feature present by calculating the ratio between the flux at an interpolated continuum level and the flux in the absorption band. A higher silicate index corresponds to a deeper silicate feature, and indicates that clouds play a more dominant role in shaping this features. The interpolated continuum is defined by two regions on either side of the absorption region. \citet{SuarezMetchev2023} shifts the continuum window at longer wavelengths to redder wavelengths, compared to \citet{SuarezMetchev2022}, to include the red wing of the silicate absorption feature, which is particularly present for young objects. However, this region at 13--14 $\mu$m has low S/N in our MIRI LRS spectra, affecting the accuracy of the interpolated continuum, and hence the calculated silicate index. The low S/N also makes it difficult to distinguish between the shape of the silicate feature at longer wavelengths. In this paper, we have used our own definition for the continuum window at longer wavelengths, which is a compromise between maintaining sensitivity to variations in the far red wing of the silicate absorption feature and sufficient S/N for the continuum. We have chosen a 0.4 $\mu$m window centered at 7.4 $\mu$m and 12.5 $\mu$m, with a linear continuum fit {(see Appendix~\ref{appendix:silice-index})}. {The 7.4 $\mu$m region lies at the end of the 6.25 $\mu$m water absorption band present in all our spectra and just before the onset of silicate absorption (Fig.~\ref{fig:opacities}), making it a pseudo-continuum region suitable for measuring the depth of the silicate feature. Although the 7.65 $\mu$m \obj{ch4} absorption feature overlaps this region, it appears only in spectra later than L8 \citep{Cushing2006,SuarezMetchev2022}, so we do not expect it to affect the silicate index measurements for our sample, except for the T1 dwarf \obj{0642}, which exhibits a strong methane feature. At this spectral type, the silicate index no longer traces the depth of silicate absorption due to the absence of that feature. However, we include silicate index values for this early T dwarf, as well as other T dwarfs in the \textit{Spitzer} sample in Fig.~\ref{fig:indices} to illustrate the overall trend of the index, noting that for T dwarfs the index is more indicative of the strength of the methane feature rather than silicate absorption.} We used this updated definition of the continuum with \verb|SEDA| to calculate our silicate indices and present them in Table~\ref{table:spectral-indices}. We applied our updated definition to the broader \textit{Spitzer} IRS sample from \citet{SuarezMetchev2022} to ensure consistency. 

{There is no 8--11 $\mu$m silicate absorption feature visible in the spectrum of \obj{0642}. These silicate clouds have sunk deep beyond the MIR photosphere, which is typical of T dwarfs \citep{AckermanMarley2001, SaumonMarley2008,Cushing2006,SuarezMetchev2021}, reinforcing our updated spectral classification.} Absorption from other molecules (eg. \obj{ch4}) in \obj{0642} change the continuum in such a way the divided spectrum does not show the silicate features, but rather the appearance of methane. The presence of methane affects the short-wavelength continuum region by reducing the flux, so the interpolated continuum is below the observed flux in the silicate absorption region. The abundance of methane increases as $T_{\text{eff}}$ decreases towards later T dwarfs, so this has less of an effect on continuum interpolation for our objects, which are earlier L-type dwarfs. However, we still expect there to be silicate clouds present in the NIR photosphere of \obj{0642} which can be indirectly observed at shorter wavelengths and can be better revealed with future retrievals. Previous atmospheric retrievals of T dwarfs have found evidence for deeper silicates \citep{Vos2023,Nasedkin2025}. \obj{0642} is the most variable at 3.6 $\mu$m in our sample, so this non-detection of the mid-IR silicate absorption feature contradicts expectations that these silicate clouds drive variability. However, different cloud compositions, such as iron, could have a greater effect on variability at NIR wavelengths.\\
\\
We compare the silicate indices of our targets to the \textit{Spitzer} IRS sample from \citet{SuarezMetchev2022} in Fig.~\ref{fig:indices}. The \textit{Spitzer} IRS sample of brown dwarfs consists of more old field dwarfs that are silicate rich than young silicate rich objects. All our targets have a measured silicate index greater than the median of the \textit{Spitzer} IRS sample, particular \obj{1741}, which has the highest silicate index yet measured. Conversely, as mentioned in Section~\ref{sec:waterandmethane}, all objects have {a} \obj{h2o} lower than the median, suggesting an anticorrelation between silicate and \obj{h2o} absorptions. {Lower gravity objects show evidence of thicker silicate clouds \citep{Marley2012,Faherty2016,Liu2016,Vos2022}, which could block} observations of water in the atmosphere \citep[][\textcolor{cobalt}{;Suárez et al. in prep}]{SuarezMetchev2023}.  Our sample consists of only young brown dwarfs, so our low \obj{h2o} indices and high silicate indices may be indicative of their youth. This can be supported by the fact that young{, low gravity} L dwarfs tend to have redder NIR colours and are more variable \citep{Kirkpatrick2008,Faherty2016,SuarezMetchev2022}, suggesting that clouds are more prominent in young L dwarf atmospheres.\\
\\
We isolated the shape of the silicate feature by dividing the flux by the interpolated continuum, as described above. \citet{SuarezMetchev2023} observed a difference in the shape of the silicate feature between young and old brown dwarfs, where their young sample had a broader and more asymmetric absorption feature that was more red-shifted compared to their old sample. This is thought to be caused by larger, heavier iron-rich silicate grains which sediment less efficiently in young objects \citep{LunaMorley2021,SuarezMetchev2023}. Similarly, we investigate the diversity in the shape of the silicate feature in our sample through comparisons with the mean silicate feature of our spectra. We present this comparison in Fig.~\ref{fig:silicate-feature}. Following a similar procedure as in \citet{SuarezMetchev2023}, we scaled the silicate feature to the mean flux in a 0.6 $\mu$m window centered at 9.3 $\mu$m. This scaling removes differences in the silicate column density, and instead highlights the diversity between the shapes, which is affected by the composition of the silicate clouds \citep[e.g.][]{LunaMorley2021,SuarezMetchev2023,Hoch2025,Molliere2025}. We also calculated the wavelength at peak absorption by fitting a third-order polynomial to the silicate feature between 8.5 -- 11 $\mu$m and estimating the wavelength at the minimum. 

From Fig.~\ref{fig:silicate-feature}, we can clearly see there is a great diversity just within our sample of L type AB Dor objects. The silicate feature of \obj{0047} and \obj{2244} are almost identical to each other, sharing similar widths and peak absorption wavelengths. These peak absorption wavelengths are relatively shifted towards longer wavelengths compared to the rest of the sample. These objects also have similarities in the absorption of other molecules, including \obj{h2o} and \obj{ch4}, as indicated by the measured molecular indices in Fig.~\ref{fig:indices} and their full spectra in Fig.~\ref{fig:opacities}. This suggests dependencies between cloud condensation and molecular abundances, including C/O ratios \citep{LunaMorley2021}. On the other hand, the silicate features of \obj{0355} and \obj{1741} are shifted to bluer wavelengths and broader compared to the other objects. \obj{1741} has similar spectral type and colour to \obj{0047} and \obj{2244}, but its silicate feature is shifted to bluer wavelengths.

\subsection{Trends with inclination}
\label{sec:trends-with-inclination}
\begin{figure}
    \centering
    \includegraphics[width=\linewidth]{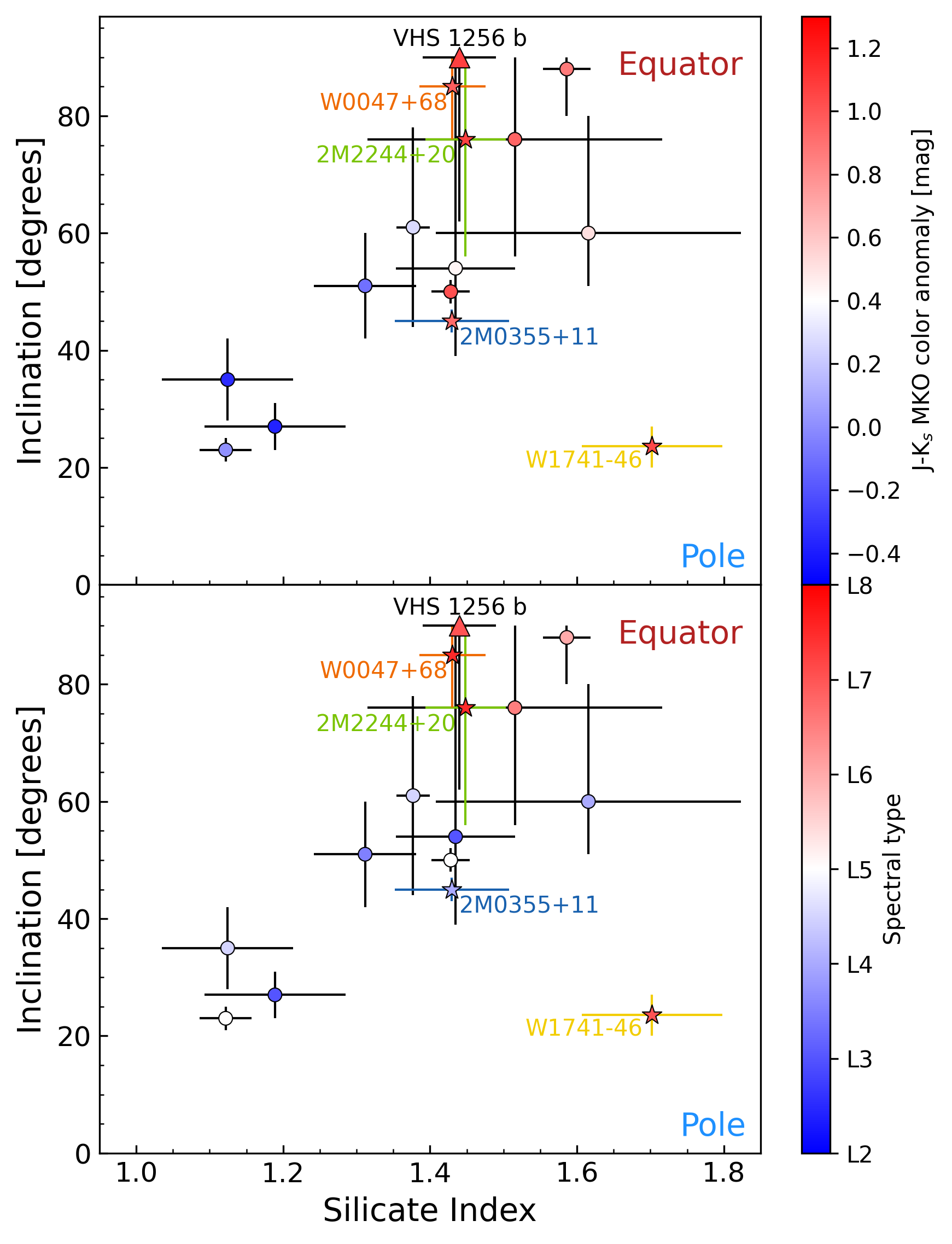}
    \caption{Correlation between the inclination and the silicate index for our sample (stars) compared with the \textit{Spitzer} sample (circles) from \protect\citet{Suarez2023}. VHS~1256~b is also included with a triangle marker and black error bars. {The upper panel is colour-coded according to} the J-K$_S$ MKO colour anomaly calculated from young objects in the UltracoolSheet \protect\citep{UltraCoolSheet}. {The bottom panel is colour-coded according to the spectral type of each object.} \protect\obj{0642} is not present in this figure as it is a T dwarf with no silicate absorption feature present in the spectrum.}
    \label{fig:silicate-inclination}
\end{figure}
\citet{Suarez2023} suggested that brown dwarfs viewed equator-on exhibited more opaque silicate clouds, while those viewed pole-on have less clouds. This also correlates with the colour anomaly, with redder objects having a more equator-on inclination angle and greater silicate index \citep{Vos2017}. In Fig.~\ref{fig:silicate-inclination}, we define our colour anomaly as the difference between the J-K$_S$ MKO colour and the mean J-K$_S$ MKO colour of all the young brown dwarfs obtained from the UltracoolSheet \citep{Best2024}. Our sample provides a unique opportunity to test the correlation between cloudiness and inclination angle, as our brown dwarfs are the same age, unlike the broader \textit{Spitzer} sample analysed in \citet{Suarez2023}. This eliminates any possibilities of cloud variations due to {surface gravity, which is similar for all objects ($\log g \sim 4.4 - 4.8$)}. Fig.~\ref{fig:silicate-inclination} shows the trend between the silicate index and inclination angle and compares it to the sample previously analysed in \citet{Suarez2023}. We did not include \obj{0642} in our analysis because its silicate index is more indicative of the depth of the methane feature than of silicate absorption. Most of our objects (\obj{0047}, \obj{2206}, \obj{2244}, and tentatively \obj{0355}) follow the previously observed trend of a strong positive correlation between inclination and silicate index. This trend suggests that lower latitude clouds have redder colours and deeper silicate features.

It is clear, however, that \obj{1741} is an outlier in our sample with an unusually high silicate index of $1.7\pm0.1$ and low inclination angle of $i={23.6^{+3.4}_{-3.6}}^{\circ}$ (viewed near pole-on). According to the correlation between inclination and silicate index, and given its nearly pole-on orientation, \obj{1741} should exhibit very weak silicate absorption. However, its MIRI spectrum reveals a strong silicate feature. \citet{TanShowman2021} suggest that latitudinal cloud distribution is determined by the object's rotation rate. For faster rotating objects ($P \leq 2$ h), we expect the cloudy equatorial bands to be more prominent compared to the clearer poles due to Coriolis effects at higher altitudes near the equator. Since \obj{1741} has a rotational period of $P\sim 15$ h, the difference between the cloud distribution at the equator and the pole is not as significant, suggesting that the whole atmosphere could be cloudy. We plan on investigating this object with atmospheric retrievals in future work to determine the cause of the unusually deep silicate feature.\\
\\
We notice a tentative trend with inclination when observing the shape of the silicate feature in Fig.~\ref{fig:silicate-feature}. The wavelength at peak absorption appears to get progressively shifted towards bluer wavelengths as the viewing angle transitions from equator-on to pole-on. The onset of the silicate absorption is also shifted towards redder wavelengths for equator-on targets. The most significant differences in the shape of the silicate feature appear to be at shorter wavelengths, although the noise at longer wavelengths in the MIRI LRS spectra make it difficult to accurately discern between each spectrum. \citet{LunaMorley2021} predicts that different grain size distributions of the same composition can produce silicate features that peak at slightly different wavelengths. More small grains will cause the absorption to be strongest at shorter wavelengths and vice versa. Our observed trend suggests that there will be more larger sized grains at the equator. In addition, they predict that grains of different compositions -- for example, enstatite and forsterite -- produce similar silicate features with only slight shifts in wavelength. Future work with atmospheric retrievals will constrain these cloud properties \citep[e.g.][]{Burningham2021,Vos2023,Molliere2025} so we can further understand how these clouds vary from equator to pole.

\section{Spectral comparisons}
\label{sec:spectral-comparisons}
\subsection{Long-term variability}
\begin{figure}
    \centering
    \includegraphics[width=\linewidth]{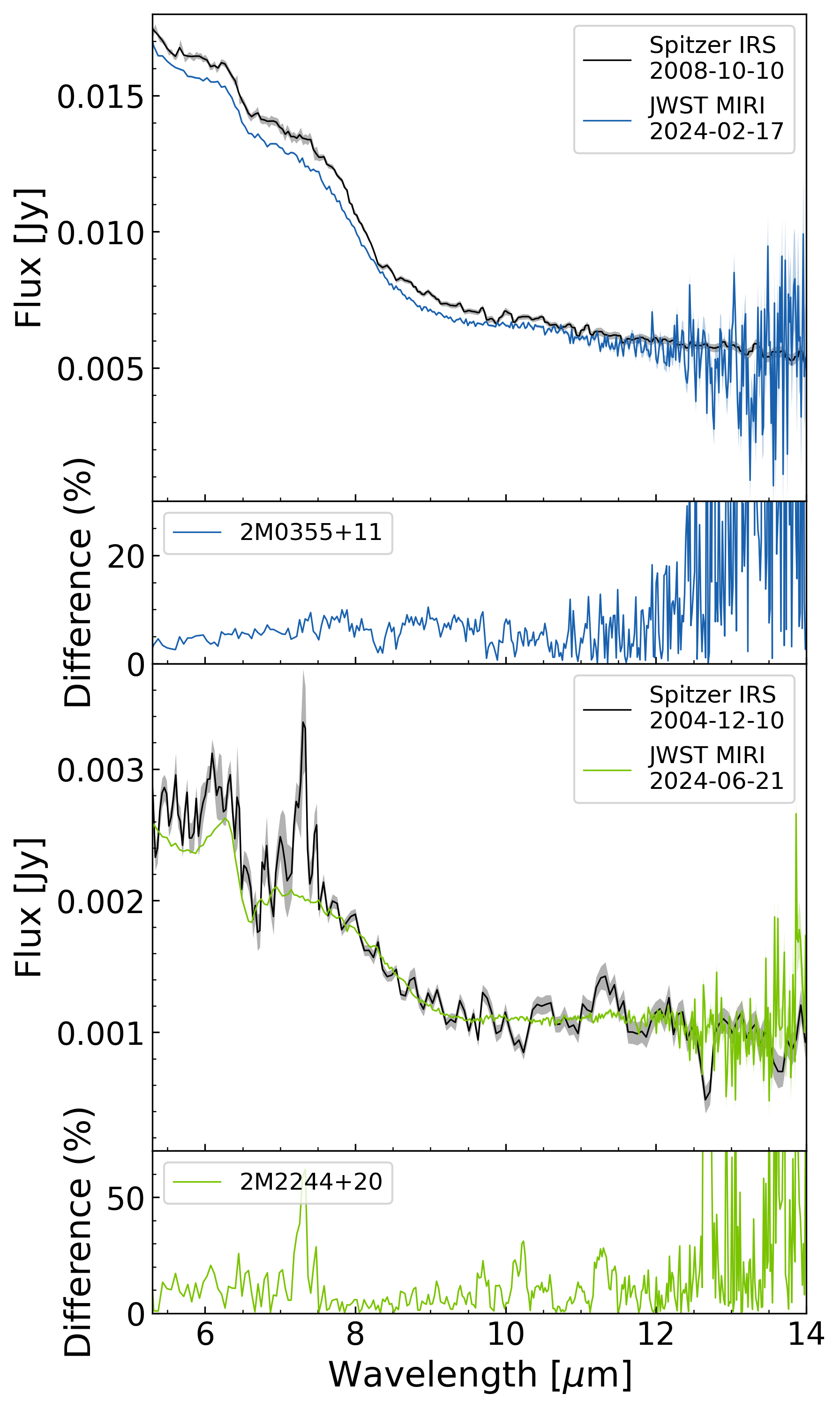}
    \caption{Spectral comparison between \textit{Spitzer} IRS observations \protect\citep{SuarezMetchev2022} and our JWST MIRI LRS observations for \protect\obj{0355} and \protect\obj{2244}. The \textit{Spitzer} spectra are shown in black and our spectra are coloured. The top two panels show the comparison for \protect\obj{0355} and the bottom two panels show the comparison for \protect\obj{2244}. The uncertainties are shown as the shaded envelope region. Below each spectral comparison is the relative flux difference (max/min $-$ 1).}
    \label{fig:spitzer-comparison}
\end{figure}

Long-term variability spanning multiple rotational periods, months or even years, has previously been detected on L/T transition dwarfs (eg. VHS~1256~b, WISE J104915.57-531906.1AB (hereafter WISE~1049AB) \citep{Luhman2013}) \citep{Apai2017,Zhou2022,Fuda2024W1049}. \citet{Zhou2022} detected long-term variability in VHS~1256~b over 1--2 years in the NIR wavelength range. VHS~1256~b is spectrally similar to our sample, and so we expect there to be similar long-term variability trends in our spectra. \citet{Chen2025} detected long-term variability across $1-12~\mu$m, including in the silicate feature, for WISE~1049AB, indicating that similar long-term variability in the silicate feature for the objects in our sample may occur.\\
\\
\citet{SuarezMetchev2022} previously analysed archival mid-IR spectra (5 -- 14 $\mu$m) for \obj{0355} and \obj{2244} with \textit{Spitzer} IRS, providing a unique opportunity to search for long-term variability. \obj{0355} was observed on 2008-10-10 and \obj{2244} was observed on 2004-12-10. We plot these \textit{Spitzer} spectra alongside our MIRI spectra in Fig.~\ref{fig:spitzer-comparison} to determine whether there is evidence for variability between epochs.

For \obj{0355} there is a flux difference of up to 10\% between the spectra at mid-MIRI wavelengths ($\sim7-10~\mu$m). However, since the spectra were observed with different instruments, there may be systematic differences that are not yet known. Because of this, we cannot be certain that the difference we observe is due to the intrinsic variability of the object or due to systematic uncertainties between \textit{Spitzer}/IRS and JWST/MIRI. Multiple epochs from JWST MIRI will allow us to determine the mid-IR variability properties of \obj{0355}.

For \obj{2244} we have good agreement between the observed spectra across the two epochs, however the \textit{Spitzer} IRS spectrum is significantly noisier. Since the spectra from both epochs are similar, we cannot identify any long-term variability in the silicate feature. Additional MIRI LRS observations, however, can more precisely measure the variability in this wavelength range without the added complication of systematics between different instruments.

The silicate index is also a measurement of variability, as silicate-rich L-dwarfs tend to be variable \citep{SuarezMetchev2022}. This measure is less sensitive to systematics as it is a relative measurement dependent on the same spectrum. Using the \textit{Spitzer} IRS spectrum, we measured the silicate index for \obj{0355} to be $1.43\pm0.03$ and for \obj{2244} to be $1.52\pm0.2$. These silicate indices agree with our measurements using JWST MIRI spectra,so we cannot confirm long-term variability in our objects. However, our high silicate indices suggest that long-term variability may exist in \obj{0355} and \obj{2244}, and we would require more time series observations to confirm.

\subsection{Comparison with directly imaged exoplanets}
\begin{figure*}
    \centering
    \includegraphics[width=\linewidth]{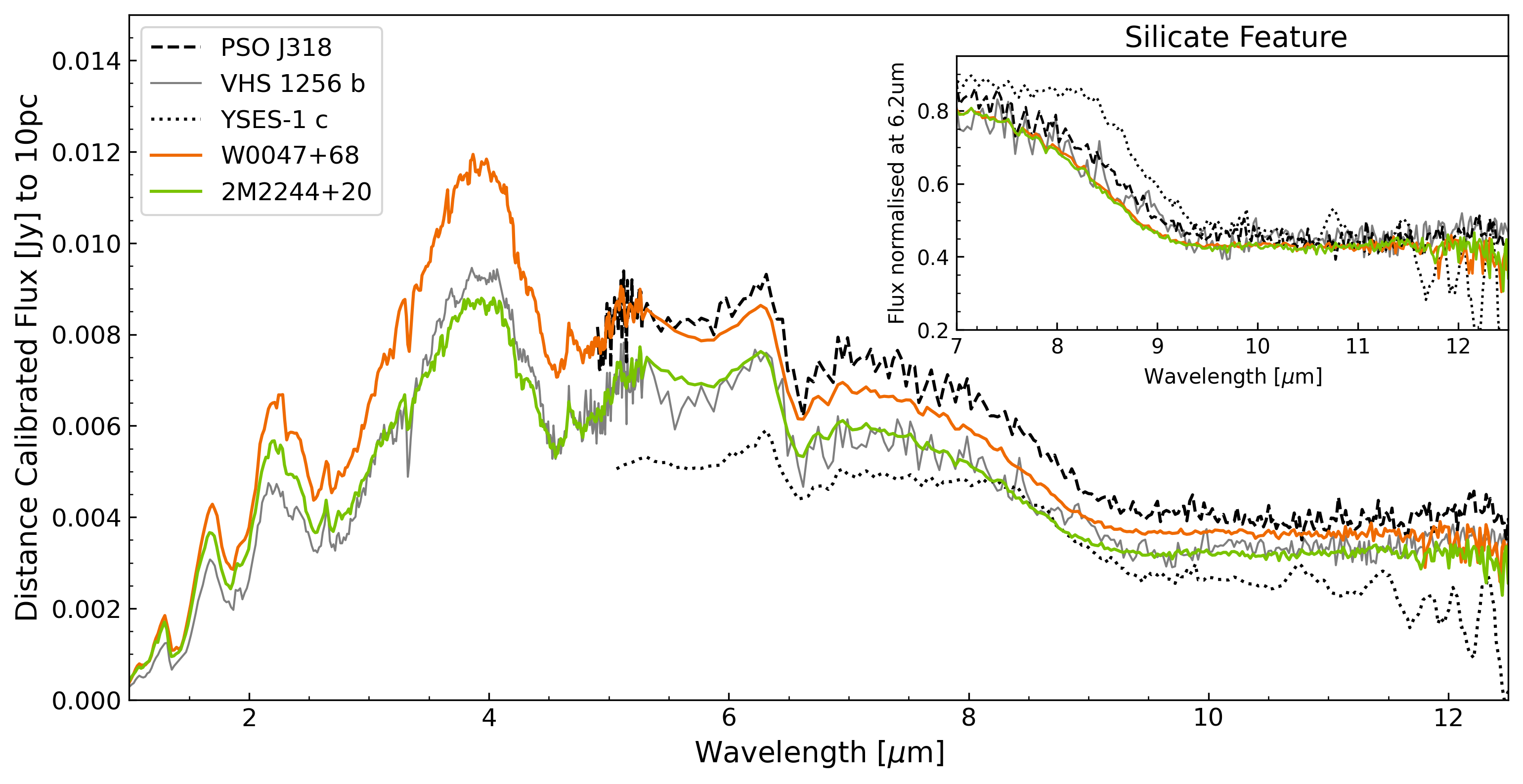}
    \caption{Comparison between our objects \protect\obj{0047} and \protect\obj{2244} (coloured lines), and other directly imaged exoplanets PSO~J318, VHS~1256~b and YSES-1~c (dashed, solid, and dotted black lines, respectively). {For ease of comparison, the spectra of each object not included in our sample were renormalised to match the spectral resolutions of our JWST NIRSpec Prism and MIRI LRS spectra. To highlight the 8 -- 11 $\mu$m silicate absorption feature,} each spectrum was normalised to the flux in a 0.2 $\mu$m window centered at 6.2 $\mu$m {and shown in the inset}. Objects in our sample are most similar to VHS~1256~b.}
    \label{fig:vhs-comparison}
\end{figure*}

Objects in our sample have similar colours and magnitudes to other previously known planetary mass objects (e.g. VHS~1256~b, PSO~J318 and YSES~1~c) (see Fig.~\ref{fig:colourmag}). We compare their spectra to examine whether there are other similarities. These objects are also similar to directly imaged exoplanets, such as HR~8799~bcd, so these comparisons will allow us to determine if our sample are good exoplanet analogues.\\
\\
VHS~1256~b {\citep{Gauza2015}} is a young \citep[$140\pm20$ Myr;][]{Dupuy2023} planetary mass companion that lies on the mass boundary between planets and brown dwarfs with an age very similar to our AB Dor objects ($\sim 133$ Myr). VHS~1256~b was the first planetary-mass object to be observed by JWST to produce a 1 -- 20 $\mu$m spectrum \citep{Miles2023}. There have been many studies on the variability and atmosphere of the most variable planetary mass object known, VHS~1256~b \citep[e.g.][]{Rich2016,Bowler2020,Zhou2022}, revealing complex cloudy atmospheres causing highly variable time-series observations.

VHS~1256~b lies along the L-T transition with spectral type L7 $\pm$ 1.5 and $\log \left(\frac{L_{\text{bol}}}{L_\odot}\right) = -4.60\pm0.05$ \citep{Miles2023}, much like our sample (see Fig.~\ref{fig:colourmag}, Table~\ref{table:fundametal-params}). Thus, this object provides an excellent scientific comparison for our sample. VHS~1256~b also shares photometric and spectroscopic similarities with other directly-imaged exoplanets like HR~8799~bcde \citep{Marois2008,Marois2010}, allowing us to view our sample as exoplanet analogues.

\obj{0047} was previously reported to be similar to VHS~1256~b \citep[e.g.][]{Zhou2022}. In our analysis, we compared the spectra of all our objects with VHS~1256~b's spectrum and plot VHS~1256~b alongside \obj{0047} and \obj{2244} in Fig.~\ref{fig:vhs-comparison}. We found that \obj{2244} has the most similar spectrum to VHS~1256~b \citep{Miles2023}, in both the shape of the absorption features and the amount of flux emitted. This is the first time that the entire SEDs have been compared, allowing us to draw novel conclusions. We downsampled the VHS 1256 b spectrum using \verb|SpectRes| \citep{Carnall2017} to match the spectral resolution achieved by NIRSpec Prism and MIRI LRS. \obj{0047} also has a similar spectrum, but is slightly more luminous than VHS~1256~b ({see Table~\ref{table:fundametal-params} and Fig.~\ref{fig:vhs-comparison}}). This observed similarity is supported by their nearby positions on the colour-magnitude diagram in Fig.~\ref{fig:colourmag}. It is notable that \obj{0047} and \obj{2244} also have mass and effective temperatures consistent within the uncertainties with VHS~1256~b \citep[$M=19\pm5 \text{ }M_{\text{Jup}}$, $T_{\text{eff}} = 1240\pm50$ K;][]{Dupuy2020}.

We have also analysed the silicate feature of VHS~1256~b. We calculated the silicate absorption features from a MIRI MRS spectrum \citep{Miles2023} that was downsampled to match our MIRI LRS resolution. Both the silicate index and the shape of the silicate feature resemble that of \obj{0047} and \obj{2244} (see Fig.~\ref{fig:vhs-comparison}). VHS~1256~b is also viewed equator-on \citep{Zhou2020} and supports the trend between inclination and silicate index (Fig.~\ref{fig:silicate-inclination}).

VHS~1256~b fits well within our sample with very similar spectra, which emphasises how the fundamental parameters of brown dwarfs, including age, temperature and mass shape their spectra. The HR8799 directly imaged exoplanet system also lies in a similar position on the colour-magnitude diagram, indicating that our sample of young AB Dor objects can be used as excellent exoplanet analogues. In order to better understand the complex atmospheric chemistry and dynamics of similar hot gas giants, we can study \obj{0047} and \obj{2244}, which we can observe much more easily than directly imaged exoplanets.\\
\\
We also compared our objects to spectra from the directly imaged isolated planetary mass object PSO~J318 \citep{Molliere2025} and the imaged companion YSES-1~c \citep{Hoch2025} in Fig.~\ref{fig:vhs-comparison}, focusing on the silicate absorption feature observed by MIRI. PSO~J318 also has a similar spectrum to \obj{0047} and \obj{2244}, emphasising our sample as analogues to planetary mass objects. \citet{Hoch2025} identified that YSES-1~c has an unusual silicate feature with a redder onset, which is made clear in the comparison in Fig.~\ref{fig:vhs-comparison}. YSES-1~c is an exoplanet viewed equator on \citep[$i\sim 90 ^{\circ}$;][]{Roberts2025}. The onset of its silicate feature shifted towards redder wavelengths supports our tentative trend between inclination and silicate absorption peak discussed in Sec.~\ref{sec:trends-with-inclination}, and suggests that this trend may also be applied to exoplanets.\\
\\
{\obj{0047} and \obj{2244} have similar magnitudes, colours, spectral types and ages to the HR 8799 planets (see Fig.~\ref{fig:colourmag}), which have dynamical masses between $6-10$ M$_{\text{Jup}}$ \citep{Brandt2021,Zurlo2022}. This indicates that these objects may have planetary masses. Dynamical masses are not possible for our isolated objects, but the evolutionary models predict a bimodal mass distribution (see Appendix~\ref{appendix:hist}), suggesting that these objects may also have planetary masses. Future studies comparing dynamical with evolutionary model masses for companions are crucial for assessing the accuracy of evolutionary models.}

\section{Conclusions}
We have presented the first full 0.6 -- 14 $\mu$m, $R\sim100$ JWST NIRSpec and MIRI spectra for a sample of five young, coeval objects in the AB Dor moving group and one young object in the Oceanus moving group. We have conducted a full SED analysis to measure the fundamental parameters of each object to fully characterise our sample. We have also analysed molecular absorption bands, in particular the silicate feature present at 8--11 $\mu$m and have investigated the possibility of latitudinal atmospheric dependencies. From this analysis, we have derived the following conclusions:
\begin{enumerate}
    \item All objects within our sample are diverse. {Our sample of AB Dor objects span from early to late L dwarfs and} we found that our objects span a range of $T_{\text{eff}} \sim 1000 - 1700$ K, $\log g \sim 4.1 - 4.8$, radius $\sim 1.20 - 1.28~\text{R}_{\text{Jup}}$ and $M \sim 8 - 34 \text{ M}_{\text{Jup}}$ (Table~\ref{table:fundametal-params}). We found that \obj{2206} and \obj{0355} were similar to each other, and also \obj{2244} and \obj{0047} were similar. {These pairs have similar spectral types to each other, so it is expected that they have similar spectra. }These two pairs could be studied together in future work to {examine} their similarities as well as the differences with respect to the full sample.
    
    \item Although \obj{0642} has the highest variability amplitude in our sample, it has no distinguishable silicate absorption feature at 10 $\mu$m. This is because at \obj{0642}'s temperature of $\sim 1070$ K and spectral type T1, the silicate clouds have most likely sunk below the photosphere. However, silicate clouds can still exist in its deeper, near-infrared photosphere and impact its NIRSpec spectrum. Future atmospheric retrievals will be able to reveal the atmospheric structure in greater detail and determine the cause of \obj{0642}'s high amplitude variability.
    
    \item \obj{1741} {is a counterexample to the suggested trend between silicate index and inclination}. This object has an unusually high silicate index ($1.7\pm0.1$) for its close to pole-on orientation (${i=23.6^{+3.4}_{-3.6}}^{\circ}$). This does not follow the expected correlation between silicate index and inclination identified in \citet{Suarez2023}. The silicate indices of the other objects in our sample agree with the trend observed in \citet{Suarez2023}. However, they have very similar silicate indices, which may be due to our selection of a uniform sample. The silicate index of \obj{1741} is also one of the highest measured in a broader sample of brown dwarfs, indicating that \obj{1741} has the most prominent silicate absorption analysed so far. {More late L dwarfs with low inclination angles are required to confirm whether \obj{1741} is an outlier or if the opposite trend is observed in later L dwarfs. This motivates future observations of late L dwarfs with low inclination angles.} In future work, we will investigate the cause of this unusual silicate absorption feature by running atmospheric retrievals with \textit{Brewster} \citep{Burningham2017}.
    
    \item We tentatively find that the shape of the silicate feature depends on inclination. Objects viewed equator-on tend to have a silicate feature shifted towards redder wavelengths, whereas those viewed closer to pole-on tend to have a silicate feature shifted towards bluer wavelengths (Fig.~\ref{fig:silicate-feature}). This could provide insights into the chemical composition and/or grain sizes of the silicate clouds, which require detailed modelling to further analyse this. We plan to further investigate whether grain size or composition are responsible for the silicate absorption differences observed in future work. This will allow us to determine the cloud structure and composition of each object and how it varies from equator-to-pole. Further analysis is also required to rule out any other factors that could be contributing to this trend (e.g. $T_{\text{eff}}$). Observations of other young objects with known inclinations (e.g. VHS~1256~b, 2MASS J05012406-0010452 \citep{Reid2008}, 2MASS J14252798-3650229 \citep{Kendall2004}, YSES$-$1~c \citep{Bohn2020}, PSO~J318) will also help to determine whether their silicate features fit within potential trends in our young sample.
    
    \item No significant long-term variability was detected in the silicate feature of two objects with literature mid-IR spectra (Fig.~\ref{fig:spitzer-comparison}). We had expected there to be a variation in the silicate feature corresponding to previously detected variability, however this was not observed in our comparisons. However, this is only from comparisons of two snapshots of spectra, and variations may still be detectable if more spectra were obtained over time. The spectra were also obtained from different instruments, so potential systematics may prevent us from probing this long-term variability.
    
    \item The objects \obj{0047} and \obj{2244} are excellent exoplanet analogues, with similarities to VHS~1256~b, another planetary mass object with similarities to known exoplanets. They also have similar spectra to the planetary mass object PSO~J318. The similarities between the spectra reinforce our motivation to study this sample of young brown dwarfs to further understand directly imaged exoplanets, such as the HR8799 system.
\end{enumerate}
 
\begin{acknowledgements}
 {We would like to thank the anonymous referee for their detailed comments and suggestions that have improved the paper.} ML and JMV acknowledge support from a Royal Society - Research Ireland University Research Fellowship (URF/1/221932, RF/ERE/221108). ML acknowledges support from Trinity College Dublin via a Trinity Research Doctoral Award. {BB acknowledges support from UK Research and Innovation Science and Technology Facilities Council [ST/X001091/1].} This work is based [in part] on observations made with the NASA/ESA/CSA James Webb Space Telescope. The data were obtained from the Mikulski Archive for Space Telescopes at the Space Telescope Science Institute, which is operated by the Association of Universities for Research in Astronomy, Inc., under NASA contract NAS 5-03127 for JWST. These observations are associated with program \#3486. This work has benefitted from The UltracoolSheet at \url{http://bit.ly/UltracoolSheet}, maintained by Will Best, Trent Dupuy, Michael Liu, Aniket Sanghi, Rob Siverd, and Zhoujian Zhang, and developed from compilations by Dupuy \& Liu (2012, ApJS, 201, 19), Dupuy \& Kraus (2013, Science, 341, 1492), Deacon et al. (2014, ApJ, 792, 119), Liu et al. (2016, ApJ, 833, 96), Best et al. (2018, ApJS, 234, 1), Best et al. (2021, AJ, 161, 42), Sanghi et al. (2023, ApJ, 959, 63), and Schneider et al. (2023, AJ, 166, 103). This work used The Immersion Grating Infrared Spectrometer (IGRINS) was developed under a collaboration between the University of Texas at Austin and the Korea Astronomy and Space Science Institute (KASI) with the financial support of the US National Science Foundation under grants AST-1229522, AST-1702267 and AST-1908892, McDonald Observatory of the University of Texas at Austin, the Korean GMT Project of KASI, the Mt. Cuba Astronomical Foundation and Gemini Observatory. The RRISA is maintained by the IGRINS Team with support from McDonald Observatory of the University of Texas at Austin and the US National Science Foundation under grant AST-1908892. This research made use of the Montreal Open Clusters and Associations (MOCA) database, operated at the Montr\'eal Plan\'etarium \citep{Gagne2026}.
\end{acknowledgements}

  \bibliographystyle{aa} 
  \bibliography{literature}
  
\onecolumn
\begin{appendix}
\section{Histograms for W0047$+$68 and 2M2244$+$20}
\label{appendix:hist}
\begin{figure*}[h]
    \includegraphics[width=\textwidth]{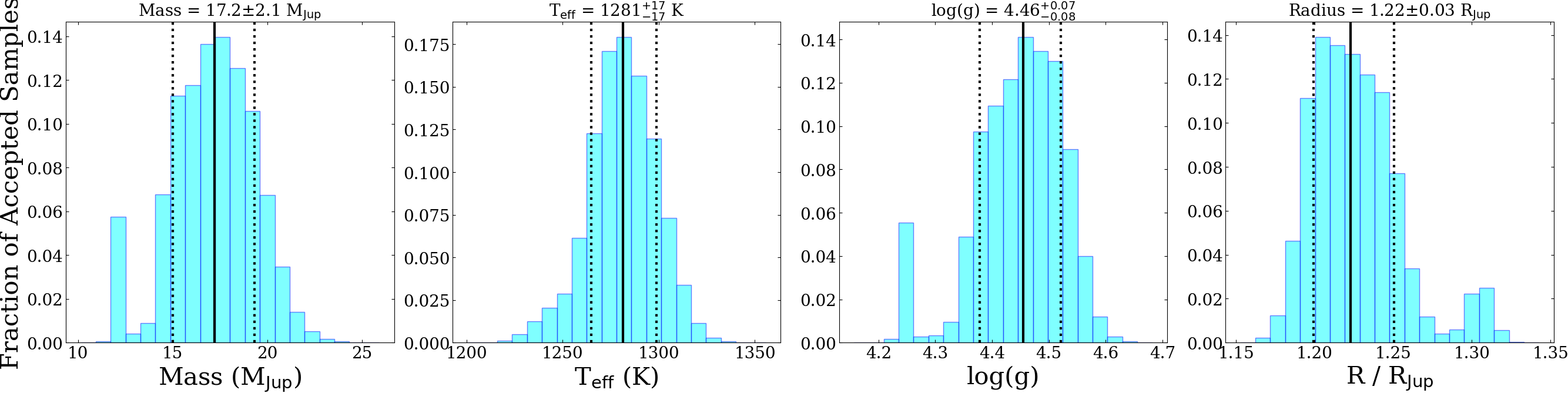}
    \caption{Histogram for fundamental parameters from rejection-sampling for \protect\obj{0047}.}
    \label{fig:0047hist}
\end{figure*}

\begin{figure*}[h]
    \includegraphics[width=\textwidth]{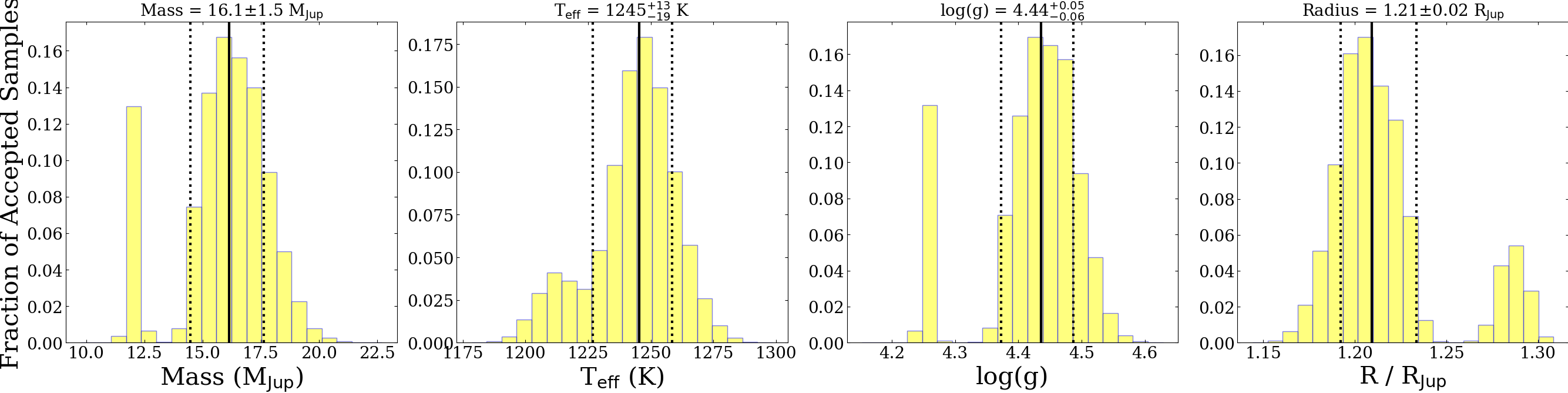}
    \caption{Histogram for fundamental parameters from rejection-sampling for \protect\obj{2244}.}
    \label{fig:2244hist}
\end{figure*}

% \twocolumn
\section{Silicate index definition}
\label{appendix:silice-index}
\begin{figure}[h]
    % \centering
    \includegraphics[width=0.5\linewidth]{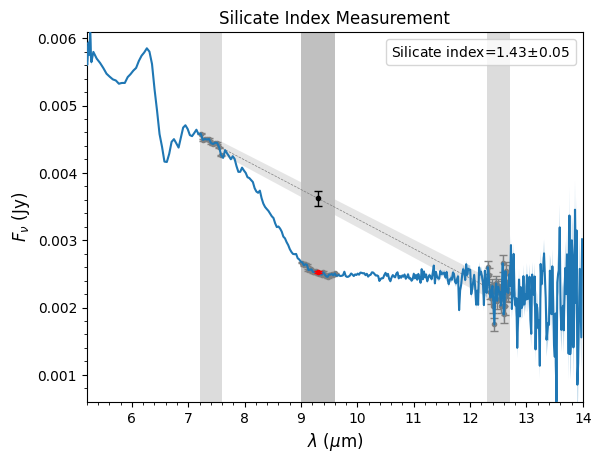}
    \caption{{Example showing the windows used to define the silicate index. \\The JWST spectrum shown here is \protect\obj{0047}. The vertical continuum \\windows are indicated in light grey. The silicate absorption window is \\highlighted in dark grey.}}
    \label{fig:silicate-index}
\end{figure}

\end{appendix}

\end{document}